 \documentclass[journal]{IEEEtran}

\ifCLASSINFOpdf
\else
\fi
\usepackage{tabstackengine}
\usepackage{cite}
\usepackage{multicol,lipsum}
\usepackage{amsmath,amssymb,amsfonts}
\usepackage{algorithmic}
\usepackage{graphicx}
\usepackage{textcomp}
\usepackage{xcolor}
\usepackage{booktabs}
\usepackage{enumerate}
\usepackage{algorithm}
\usepackage{multirow}
\usepackage{standalone}
\usepackage{tikz}
\usetikzlibrary{positioning}
\usetikzlibrary{decorations.markings}
\usepackage{float}
 \usepackage{makecell}
\usepackage{subcaption}
\usepackage{arydshln}
\usepackage{url}

\makeatletter
\newcommand{\Biggg}{\bBigg@{4}}
\newcommand{\Bigggg}{\bBigg@{5}}
\makeatother

\begin{document}

\title{On the Design of Variable Modulation and Adaptive Modulation for Uplink Sparse Code Multiple Access}
%

\author{ Qu Luo,  \textit{Member, IEEE},  Pei Xiao, \textit{Senior Member, IEEE},  Gaojie Chen, \textit{Senior Member, IEEE}, and Jing Zhu, \textit{Member, IEEE}.

\thanks{

 Q. Luo,  P. Xiao  and J. Zhu   are with 5G and 6G Innovation centre, Institute for Communication Systems (ICS) of University of Surrey, Guildford, GU2 7XH, U.K.  (e-mail: \{q.u.luo, p.xiao, j.zhu\}@surrey.ac.uk). 
 
 Gaojie Chen is with the School of Flexible Electronics (SoFE) \& State Key Laboratory of Optoelectronic Materials and Technologies (OEMT), Sun Yat-sen University, Shenzhen, Guangdong 518107, China (e-mail: gaojie.chen@ieee.org).
}
}
 
%
 

\maketitle
\pagestyle{empty}
\thispagestyle{empty}

\begin{abstract}
Sparse code multiple access (SCMA) is a promising  non-orthogonal multiple access scheme for   enabling   massive connectivity  in next generation wireless networks.  However, current SCMA codebooks are  designed with the same size, leading to inflexibility of user grouping and supporting diverse data rates.  To address this issue, we propose a variable modulation SCMA (VM-SCMA)  that allows users to employ codebooks with different modulation orders.  To guide the VM-SCMA design, a VM matrix (VMM) that assigns modulation orders based on the SCMA factor graph  is first introduced.  We formulate the VM-SCMA design using the proposed average inverse product distance and the asymptotic upper bound of sum-rate, and  jointly optimize the  VMM, VM codebooks, power  and codebook allocations.  The proposed VM-SCMA not only enables diverse date rates  but also supports different  modulation order combinations  for each rate.  Leveraging these distinct advantages, we further  propose an adaptive VM-SCMA (AVM-SCMA) scheme which adaptively selects the rate and the corresponding VM codebooks to adapt to the users' channel conditions by maximizing the proposed effective throughput. Simulation results show that the overall designs are able to simultaneously achieve a high-level  system flexibility, enhanced error rate  results, and  significantly improved throughput performance, when compared to conventional  SCMA schemes.

\end{abstract}


%
\IEEEpeerreviewmaketitle

\section{Introduction}
%
%
%
%
\IEEEPARstart{N}{on-orthogonal} multiple access (NOMA) has been envisioned as a promising multiple access technique for future wireless networks,  such as  the   beyond fifth generation (B5G) and sixth-generation (6G) networks,  to address the increasing demand for high spectral efficiency and massive connectivity \cite{NGMA1, NGMA2, YuSparseStandards}. Unlike the conventional orthogonal multiple access (OMA) technologies,  such as time division multiple access (TDMA) and orthogonal frequency division multiple access (OFDMA), NOMA users are allowed communicate  simultaneously  over  the same radio resources, i.e.,   time and frequency resources \cite{hoshyar2008novel}. Sparse code  multiple access (SCMA),   combines both modulation and spreading procedures,  has been considered as a  key candidate of the code-domain NOMA (CD-NOMA), and consequently has attracted a sustained research attention from both   academia and industry \cite{SSD1,TaherzadehSCMA,liu2021sparse,LuoMultitask,LuoError}.   Based on the well designed  codebooks, the incoming message bits from SCMA users are directly mapped to multi-dimensional codewords \cite{TaherzadehSCMA},  which are intentionally  sparse so as to reduce the  decoding complexity by employing the  message passing algorithm (MPA).    

\vspace{-0.3cm}
\subsection{Related Works}  
Designing efficient sparse codebooks is considered as a fundamental approach to enhance the spectrum efficiency in SCMA.  
 In \cite{TaherzadehSCMA}, a multi-stage process for efficient codebook design  was developed,     consisting of the mother constellation (MC)   and the user-specific constellations (e.g., shuffling, rotation angles) designs. In 
 \cite{YuDesign}, the Star-QAM (quadrature amplitude modulation) signal constellation was employed  to design the MC. In contrast,  the golden angle modulated   signals were utilized to design the MC with lower peak-to-average power ratio in \cite{mheich2018design}.  Near-optimal codebook designs for SCMA schemes were investigated for different channel environments in \cite{chen2020design}.  Unlike the above works where  the minimum Euclidean distance (MED) or  minimum product distance (MPD) is employed to aid the codebook design, the codebooks reported in \cite{XiaoCapacity} and \cite{JiangLow} were obtained by maximizing the constellation constrained input channel capacity and  mutual information, respectively. More recently,  a novel class of low-projection SCMA codebooks for ultra-low decoding complexity in downlink Ricain fading channels were proposed  in \cite{LPCSMA}.

It is noted that the above works  mainly design SCMA codebooks based on a regular structure with same codebook size. Nevertheless, this implicitly assumes that   the SCMA users   sharing the same radio resources have a similar quality of service (QoS)  requirement  and channel condition \cite{CheraghyJoint}, which may not be realistic in  practical scenarios.  To  meet the diverse  QoS  requirements of SCMA users,  an alternative approach is to  perform user grouping and resource allocation at the base station (BS) \cite{CheraghyJoint,Jaber1,Evangelista}, where a sum-rate   is generally considered as the optimization metric. However, it is noted that the achievable sum-rate of SCMA  systems relies heavily on the sparse codebooks \cite{CheraghyJoint}. Moreover, when the  QoS requirement or channel condition of a SCMA user changes, the  BS needs to re-allocate resources or re-group the users if the codebook with same codebook size is employed,  which  can  result in excessive signaling  overhead  and unaffordable complexity at the BS.  In a nutshell, current SCMA codebooks with the same modulation order pose a challenge for   user-grouping and  supporting  diverse data rates. 





Adaptive modulation (AM) is an effective method   for enhancing the spectral efficiency of SCMA by dynamically selecting the codebook with a suitable size (modulation order)  based on the users' channel conditions or service needs.
  The AM technique has been widely applied to wireless communication systems \cite{SeongDegrees}. A systematic study on the improvement in spectral efficiency obtained by optimally varying combinations of the modulation order, power, and instantaneous bit-error rate (BER) was conducted in \cite{SeongDegrees}. Recently, the AM schemes have been extensively studied in power-domain NOMA (PD-NOMA) systems  \cite{YahyaAdaptive, YuJoint,WangAsymmetric}. For example,  the joint adaptive $M$-quadrature amplitude modulation ($M$-QAM)  and power adaptation for   a downlink two-user PD-NOMA network was investigated in \cite{YuJoint}. In contrast,   an asymmetric AM framework is introduced  in \cite{WangAsymmetric} for uplink PD-NOMA systems, where the closed-form BER expressions were derived. Despite the widespread investigation of SCMA in the literature,  to the best of the authors’ knowledge, very little work  has  considered the spectral efficiency maximization problem for SCMA systems using AM technique.

\vspace{-0.3cm}

\subsection{Motivations and Contributions }

 Despite the extensive   literature  on SCMA system designs, a few core issues have yet to be properly addressed.  Firstly, existing SCMA systems    treat all users equally and employ the codebook with the same modulation order,  leading to inflexibility of supporting diverse data rates.  Hence, it is  pivotal to design a class of more flexible SCMA codebooks that enable diverse modulation orders  whilst improving the error rate performance.   Secondly, most  resource management schemes on SCMA are based on Shannon theory, without placing any restrictions on the practical codebooks.    Thirdly,  the amalgamation of SCMA and AM is considered as a disruptive approach to further facilitate the system spectral efficiency.  It is, however, challenging to investigate an adaptive codebook allocation scheme since the   performance of  the SCMA users are mutually coupled.

 To tackle the above issues,  we first propose a novel SCMA scheme, called variable modulation SCMA (VM-SCMA) where  SCMA users are allowed to employ codebooks with different codebook sizes (modulation orders).  
  Hence, a fundamental problem in VM-SCMA is how  to design efficient  VM sparse  codebooks. It is noted that the traditional  MPD and MED based  codebook design metrics may not be applicable  since they are derived based on   regular structure with the same modulation order \cite{YuDesign,mheich2018design,chen2020design,LPCSMA,ZhangUniquely}.  In this paper, the   sum-rate and pair-wise  error probability (PEP) are employed  to guide the VM codebook design.    Considering the  fairness of different users, we propose to minimize the   error rate performance of the worst user. In addition, we consider a more realistic scenario   where users are  located at different locations within a cell. Accordingly, the  near-far effect is taken into account in the design of VM-SCMA. Since the proposed VM-SCMA enables different modulation orders for a given overall date rate $R_b$,  a fundamental problem naturally arises, i.e.,  how to adaptively select the $R_b$ and  the optimal VM codebooks according to the users' statistical channel conditions.  To this end, we propose to maximize the  effective throughput of   VM-SCMA systems subject to a reliable error rate constraint, which is referred to as adaptive  VM-SCMA (AVM-SCMA) system.

To summarize, the  main contributions of this work are as follows:

\begin{itemize}
    \item We propose a novel uplink VM-SCMA system with randomly deployed users,  which serves as a general framework and subsumes the existing SCMA systems as special cases.  The average inverse product distance (AIPD) of a codebook  and a symbiotic sum rate upper bound are derived to guide the VM-SCMA design.
    
    \item We develop a VM matrix (VMM) that involves the  modulation order assignment  based on a factor graph  to facilitate the VM codebook design. Moreover, low complexity solutions are developed  for the design of VMM, near-optimal MCs, power allocation and codebook allocation.
    
  \item To further enhance spectrum efficiency, we propose an  AVM-SCMA, which    adaptively chooses the  VM codebooks based on users' statistical signal-to-noise ratios  (SNRs).   An error rate performance model based on the received  statistical SNRs is proposed to reduce the prohibitively high computational complexity of calculating the effective throughput of VM-SCMA.

   \item We  conduct extensive numerical experiments to show the superiority of the proposed VM-SCMA and AVM-SCMA schemes.  The simulations demonstrate that the overall
designs are able to simultaneously provide a high-level flexibility of supporting diverse data rates, and significantly improve error rate performance and throughput  compared to conventional SCMA schemes.

\end{itemize}

\vspace{-0.3cm}
     \subsection{Organization}

The remainder of this paper is organized as follows.
Section II presents the SCMA system model with randomly deployed users. Section III
presents the basic concepts of the proposed VM-SCMA. The sum-rate and error rate performance of VM-SCMA are investigated. Accordingly, the design of VM-SCMA is formulated. Next, the detailed design of VM-SCMA is 
presented in Section IV.  Section V  introduces the proposed  AVM-SCMA by maximizing the effective throughput.  Numerical results are presented in Section VI to validate the performance the proposed algorithms. Finally, concluding remarks are drawn in Section VII. 

\vspace{-0.3cm}
           \subsection{Notation}

 The $n$-dimensional complex and binary vector spaces are denoted as $\mathbb{C}^n$ and $\mathbb{B}^n$, respectively. Similarly, $\mathbb{C}^{k\times n}$ and $\mathbb{B}^{k\times n}$ denote the $(k\times n)$-dimensional complex and binary matrix spaces, respectively. ${{\mathbf{I}}_{n}}$ denotes an $n \times n $-dimensional  identity matrix. $\text{tr}(\mathbf{X})$ denotes the trace of a square matrix $\mathbf{X}$.   $\text{diag}(\mathbf{x})$ gives a diagonal matrix with the diagonal vector of $\mathbf{x}$. $(\cdot)^\mathcal T$, $(\cdot)^ \dag $ ,  $(\cdot)^\mathcal H$  and $\mathbb E(\cdot)$ denote the transpose, the conjugate, the Hermitian transpose, and expectation operation, respectively.  $  \mathcal {CN}(0,1)$ denotes the standard Gaussian  distribution with zero mean and unit variance. $\|\mathbf{x}\|_2$ and $|x|$ return the Euclidean norm of vector $\mathbf{x}$ and the absolute value of $x$, respectively.

\section{System model}

 \begin{figure}[htbp]
  \centering
  \includegraphics[width= 3.5in]{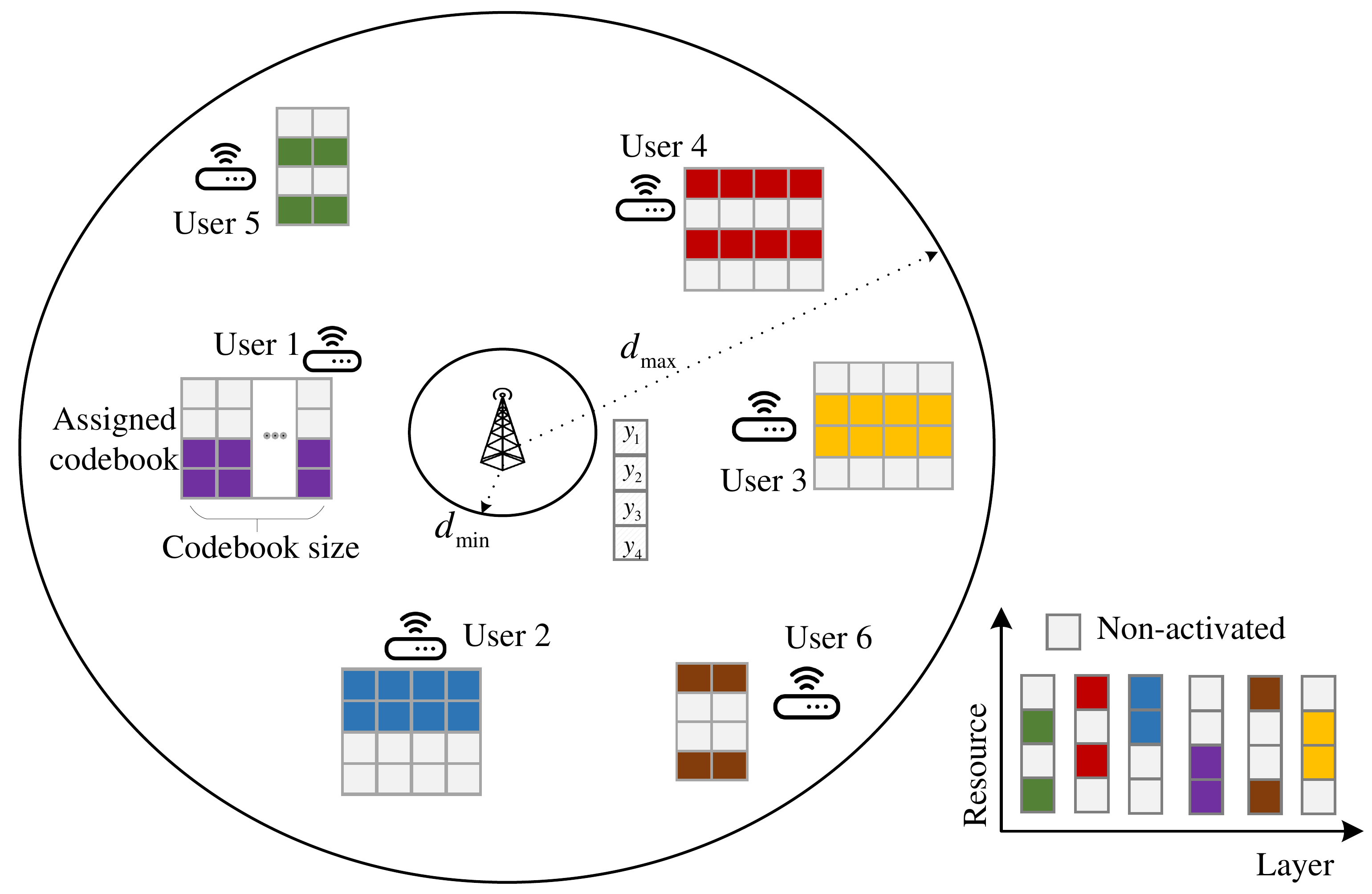} 
  \caption{Uplink  SCMA system with randomly deployed users.}
\label{sys}
\vspace{-0.3cm}
\end{figure}

\begin{figure}
  \centering
  \includegraphics[width=3.2in]{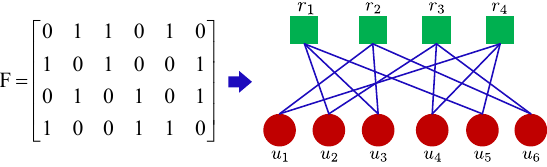} 
  \caption{Factor representation of a SCMA system.}
\label{factor}
\vspace{-0.3cm}
\end{figure}

We consider an uplink  SCMA system   consisting  of  a  BS  and  $J$ randomly distributed users   served by   $L$   codebooks, each consisting of $K$ orthogonal resource  nodes (RNs). The overloading factor, defined     as $ \lambda = \frac{L}{K}$, is   larger than $100\%$. Let      $\boldsymbol {\mathcal X} = \left\{\boldsymbol {\mathcal X}_{1},\boldsymbol {\mathcal X}_{2}, \ldots, \boldsymbol {\mathcal X}_{L} \right\} $  be the set of $L$ codebooks, where $ \boldsymbol {\mathcal X}_{l}   \in \mathbb {C}^{K \times M_l}$ denotes the $l$th codebook with   modulation order (codebook size) of  $M_l$.  Fig. \ref{sys} shows an example of uplink SCMA system with $K=4$ and $L =6$.
 The $K$-dimensional complex codewords in  $\boldsymbol {\mathcal X}_{l}, l= 1,2,\ldots, L $ are sparse vectors with $N$ non-zero elements and $N < K$.   The sparse structure of   $\boldsymbol {\mathcal X}$ can be represented by an indicator (factor graph) matrix  $\mathbf {F}_{K \times L} = \left [ \mathbf {f}_1, \ldots, \mathbf {f}_L \right] \subset \mathbb {B}^{K\times L}$.  An  element in ${\bf{F}}$ is defined as ${f_{k,l}}$, and the variable node (VN)  $v_l$ is connected to  the   RN     $r_k$ if and only if $f_{k,l}=1$.  A factor graph can graphically represents the sharing of the  RNs among multiple   VNs.   Fig. \ref{factor} illustrates an example of the indicator matrix and the corresponding factor representation of $K=4, J=6, d_v =2$ and $d_f =3$.

 We further denote  the binary matrix $\mathbf W = \left\{ w_{l,j} \right\}^{L \times J}  $ as the codebook allocation matrix, where $w_{l,j} \in \{0,1\}$ and $w_{l,j}=1$ denotes the $l$th codebook is allocated to the $j$th user. In this paper, we assume $L =J$ and each user is assigned with  only one  codebook, i.e., $\sum \nolimits_{l=1}^{L} w_{l,j} =1, j= 1,2,\ldots,J $. In the rest of the paper, we denote  $l$ and $j$ as the   subscripts  for the codebook and user  indexes, respectively, and employ $J$ to represent the total number for codebooks and users.
Then, the $j$th user's codebook is given by $\boldsymbol {\mathcal S}_{j} = \sum \nolimits_{l=1}^{J} w_{l,j} \boldsymbol {\mathcal X}_{l}$. Accordingly, the sparse structure of $\boldsymbol {\mathcal S} = \left\{\boldsymbol {\mathcal S}_{1},\boldsymbol {\mathcal S}_{2}, \ldots, \boldsymbol {\mathcal S}_{J} \right\} $ can be represented by an  equivalent indicator matrix,  which is given by $ \widetilde{\mathbf F}  = \left [ \widetilde{\mathbf {f}}_1,\widetilde{\mathbf {f}}_2, \ldots, \widetilde{\mathbf {f}}_J \right]= \mathbf F \mathbf W $. The set of  column indices sharing the $k$th RN  of ${\mathbf F}$ and $\widetilde{\mathbf F}$ are defined as  $  \boldsymbol{\boldsymbol{\phi}}_{\mathbf F}(k)=
\left\{ {j\left| {{f_{k,j}} = 1} \right.} \right\}, $ and $   \boldsymbol{\boldsymbol{\phi}}_{\widetilde{\mathbf F}}(k)=
\left\{ {j\left| {{\widetilde{f}_{k,j}} = 1} \right.} \right\}$, respectively. 
  Similarly,   define $\boldsymbol{\boldsymbol{\varphi}}_{\mathbf F}(j)=
\left\{ {k\left| {{f_{k,j}} = 1} \right.} \right\},    \boldsymbol{\boldsymbol{\varphi}}_{\widetilde{\mathbf F}}(j)=
\left\{ {k\left| {{\widetilde{f}_{k,j}} = 1} \right. } \right\}$ as the set of row indices sharing the  $j$th VN  of ${\mathbf F}$ and $\widetilde{\mathbf F}$, respectively. 

Let $\mathbf {h}_{j}=[ {h}_{j,1}, {h}_{j,2}\ldots , {h}_{j,K}]^{{ \mathcal T}}$ be the channel vector between the $j$th user to the BS, where $ {h}_{j,k}= {g}_{j,k} d_{j}^{-\frac {\alpha }{2}}$,  $g_{j,k} \sim \mathcal {CN}(0,1)$ denotes   a Rayleigh fading coefficient, $d_j$ is the distance between the BS and the $j$th user, and $\alpha \geq 1$ is the path loss coefficient.  Let $\mathbf d= [d_1, d_2, \ldots, d_J]^{\mathcal T}$ be the distance vector of $J$ users. Without loss of generality,   we assume $d_1  \geq  d_2\geq  \ldots, \geq  d_J $. The path loss model is given  by \cite{JointCheraghy}
\begin{equation}
\small
     {d_{j}}^{-\alpha/2 }=\begin{cases}~ {d_{j}}^{-\alpha/2 },&  {d_{\min}} <  {d_{j}}\leq  {d_{\max}},\\  {d_{\min}}^{-\alpha/2 }, &   {d_{j}} \leq d_{\min}, \end{cases}
\end{equation}
where $d_{\min}$ and $d_{\max}$ are the minimum and maximum user link distances, respectively. The $K$-dimensional received signal $\mathbf y $ at the BS is expressed as 
  \begin{equation}  
  \label{UPlink}
  \small
  \mathbf {y}=\sum \limits _{j=1}^{J}  \text {diag}(\mathbf {h}_{j})\sqrt {p_{j}}\mathbf {s}_{j} +\mathbf {n},
 \end{equation}
where $\mathbf {s}_{j} $ is the transmitted codeword  of the $j$th user drawn from  $ \boldsymbol {\mathcal S}_{j} $, $\mathbf {n}=[n_{1},n_{2},\ldots,n_{K}]^{\mathcal  T}$ denotes the Gaussian noise (AWGN) vector with $n_{k} \sim \mathcal {CN}(0,N_{0})$  and  $p_j, 1 \leq j\leq J$,  is the transmit  power of user $j$.

\vspace{-0.3cm}

 \subsection{MPA}  
   Benefiting from the sparsity of SCMA codewords,  the MPA detector is  applied to reduce the decoding complexity while achieving maximum likelihood performance.  The MPA algorithm employs an iterative approach for multi-user detection, wherein it calculates probability distributions for edges using a factor graph structure  \cite{Chaturvedi}. Denote $I_{{{{r}_{k}}\to {{v}_{j}}}^{{}}}^{(t)} ({\mathbf{x}_{j}})$ and $I_{{{v}_{j}}\to {{r}_{k}}}^{{(t)}} ({\mathbf{x}_{j}})$ as the belief message propagated from ${r}_{k}$  to   ${v}_{j}$ and ${v}_{j}$   to  ${r}_{k}$ at the $t$th iteration, respectively.  The iterative messages exchanged between  RNs and  VNs are computed as
 \begin{equation}
  \small
  \label{FN_up}
 \begin{aligned}
 I_{r_k \rightarrow v_j}^{(t)}(\mathbf {x}_j) & \\ = \sum _{\substack{ i\in  \boldsymbol{\boldsymbol{\phi}}_{\widetilde{\mathbf F}}(k) \backslash \lbrace j\rbrace \\ \mathbf {x}_i \in \mathcal {X}_{i}}} &\frac{1}{\sqrt{2\pi N_0^2}} \text{exp} \left\{-\frac{ \vert y_k -  h_{k}  \sum  _{i\in  \boldsymbol{\boldsymbol{\phi}}_{\widetilde{\mathbf F}}(k)}x_{k,i} \vert ^2}{2N_0} \right\}\\ &\times \prod _{i\in  \boldsymbol{\boldsymbol{\phi}}_{\widetilde{\mathbf F}}(k) \backslash \lbrace j\rbrace } I_{v_i \rightarrow r_k}^{(t -1)}(\mathbf {x}_i), 
 \end{aligned}
\end{equation}
and
\begin{equation}
 \small
  \label{VN_up}
I_{v_j \rightarrow r_k}^{(t)}(\mathbf {x}_j) =\kappa _j\times \prod _{\ell \in \boldsymbol{\boldsymbol{\varphi}}_{\widetilde{\mathbf F}}(j) \backslash \lbrace k\rbrace } I_{r_{\ell } \rightarrow  v_j}^{(t-1)}(\mathbf {x}_j), \end{equation}
where $\kappa _j$ is  a normalization factor, $\boldsymbol{\boldsymbol{\phi}}_{\widetilde{\mathbf F}}(k) \backslash \lbrace j\rbrace $ represents removing the $j$th VN from the set
$\boldsymbol{\boldsymbol{\phi}}_{\widetilde{\mathbf F}}(k)$,  and $\boldsymbol{\boldsymbol{\varphi}}_{\widetilde{\mathbf F}}(j) \backslash \lbrace k\rbrace$ represents removing the $k$th RN from the set  $\boldsymbol{\boldsymbol{\varphi}}_{\widetilde{\mathbf F}}(j)$.  The prior probability of the codeword is
initialized as
\begin{equation}
    \small
    I_{{{r}_{k}}\to {{v}_{j}}}^{(0)}({\mathbf{x}_{j}})=\frac{1}{M_j},   \forall j=1,  \ldots, J, \forall k \in \boldsymbol{\boldsymbol{\varphi}}_{\widetilde{\mathbf F}}(j).
\end{equation}

 \section{The Proposed VM-SCMA}
 \label{VM-SCMA}

We  consider a generalized case,  where    $J$ SCMA users are randomly distributed   within a cell and   can  employ  the sparse codebooks  with  different modulation  orders.  To better   demonstrate the idea of  VM-SCMA, we   first define  $\boldsymbol {\mathfrak{m}} =\left\{M_{l'} \right\} ^{J \times 1}$ as   the set of possible    modulation order combinations  that achieve  a  total rate of  $R_b$, i.e., $\sum\nolimits_{l'=1}^{J} \log_2({M_{l'} })=R_b$.   Then, we   introduce the following definition:

 \textit{Definition 1:} 
Define  a  \textit{VMM}, denoted by  ${{\mathbf{M}}_{K\times J}}$, as the allocated modulation orders from  $\boldsymbol {\mathfrak{m}}$  to the $J$ codebooks    based on the factor graph. Considering the example where $K = 4,  J = 6, d_v = 2,$ and $d_f = 3$, the VMM is defined as
  \begin{equation} 
 \label{Model2}
 \small
 {{\mathbf{M}}_{4\times 6}}=\left[ \begin{matrix}
  0 & M_2 & M_3 & 0 &  M_5 & 0  \\
  M_1 & 0 & M_3 & 0 & 0 & M_6\\
  0 & M_2 & 0 & M_4 & 0 & M_6 \\
  M_1 & 0 & 0 & M_4 & M_5 & 0  \\
\end{matrix} \right], \forall M_l \in  \boldsymbol {\mathfrak{m}},
  \end{equation}
where $ M_l $ is the assigned modulation order for the $l$th layer. Note that we will frequently use a VMM in Section \ref{Design}. 

Clearly, for a given valid  $R_b $, there exist   multiple $\boldsymbol {\mathfrak{m}}$, and for each $\boldsymbol {\mathfrak{m}} $, there are  $J !$   allocation strategies which can lead  to different VMMs. In existing SCMA codebook designs, $M_{l'} = M_l, 1 \leq \forall l',l \leq J $ is generally assumed. In contrast,  the proposed VM-SCMA supports different modulation orders for the   $J$ codebooks.

After define the AMM, we present  the major differences between the proposed VM-SCMA and AVM-SCMA in  Remark 1. 

\textit{Remark 1:   In the proposed VM-SCMA, we aim to design the optimal VM codebooks including the VMM (i.e., ${{\mathbf{M}}_{4\times 6}}$), codebook allocation  and  power allocation under a given $R_b$. For example, for a given rate of  $R_b = 12$ bits, one can choose   different modulation order combinations, e.g.,  $\boldsymbol {\mathfrak{m}} = \{4,4,4,4,4,4\}$ and $\boldsymbol  {\mathfrak{m}} = \{2,2,2,4,8,16\}$. In addition, for each $\mathfrak{m}$, the ${{\mathbf{M}}_{4\times 6}}$ can also be different. For example, the following  are two valid VMMs for $\boldsymbol  {\mathfrak{m}} = \{2,2,2,4,8,16\}$:
     \begin{equation} 
           \small    
 \label{Model1}
       \setlength{\arraycolsep}{1.5pt} 
 \begin{aligned}
 &{{\mathbf{M}} _{4\times 6}^{(1)}}=\left[ \begin{matrix}
  0 & 2 & 2 & 0 &  8 & 0  \\
  2 & 0 & 2 & 0 & 0 & 16\\
  0 & 2 & 0 & 4 & 0 & 16 \\
  2 & 0 & 0 & 4 & 8 & 0  \\
\end{matrix} \right],
{{\mathbf{M}}_{4\times 6}^{(2)}}=\left[ \begin{matrix}
  0 & 2 & 2 & 0 &  2 & 0  \\
  4 & 0 & 2 & 0 & 0 & 16\\
  0 & 2  & 0 & 8 & 0 & 16 \\
  4 & 0  & 0 & 8 & 2 & 0  \\
\end{matrix} \right].
 \end{aligned}
       \end{equation} 
 Hence, the goal of  VM-SCMA design is to determine the optimal VMM, codebook and power allocation under the data rate constraint $R_b$.}

\textit{Building upon the proposed VM codebooks, the AMV-SCMA aims to adaptively choose a large   $R_b$ based on users' statistical  channel conditions under a reliable error rate constraint. The detailed design of AVM will be presented in Section  \ref{AVMSec}. }
 


 \subsection{Fundamentals of VM-SCMA}

\label{Fundamentals}
 
\textit{1) Sum rate of VM-SCMA:} 
The signal model of (\ref{UPlink}) can be written in an equivalent MIMO channel model, i.e.,
 \begin{equation}
 \label{MIMOmodel}
     \mathbf {y} = \mathbf {H} \mathbf {s} + \mathbf {n},
 \end{equation}
 where $\mathbf {H} = [\mathbf {H}_{1}, \mathbf {H}_{2}, \cdots, \mathbf {H}_{J}] \in \mathbb C^{K \times KJ}$, $\mathbf {H}_{j} = \text {diag}(\mathbf {h}_{j})$, and   $\mathbf {s} = \left [ \sqrt{p_1} \mathbf  {s}_{1}^{\mathcal T},  \sqrt{p_2}\mathbf {s}_{2}^{\mathcal T}, \ldots, \sqrt{p_J}\mathbf {s}_{J}^{\mathcal T} \right]^{\mathcal T}$.  Denote the covariance matrix of $\mathbf s$   by $\mathbf {K}_{s}   \in \mathbb {C}^{KJ \times KJ}$. 
Based on the MIMO signal model in (\ref{MIMOmodel}), the maximum mutual information between the transmitted signal   $\mathbf {S} = \left [ \mathbf {s}_{1},  \mathbf {s}_{2}, \ldots, \mathbf {s}_{J} \right]$  and received signal is given by \cite{tse2005fundamentals}
 \begin{equation} 
 \small
 \label{AMI}
 \begin{split}  &I(\mathbf{S};\mathbf{y}\vert\boldsymbol{H}) =  \log\det(\boldsymbol{I}_{K}+\frac{1}{N_{0}}\boldsymbol{HK}_{s}\boldsymbol{H}^{\ast})\\ &=  \log\det\begin{pmatrix} 1+\frac{\sum  \limits_{j=1}^{J} h_{j,1}^{2}p_j \mathbb E(s_{j,1}^{2})}{N_{0}} & \cdots & 0\\ \vdots & \ddots & \vdots\\ 0 & \cdots & 1+\frac{\sum \limits_{j=1}^{J}h_{j,k}^{2}p_j \mathbb E(s_{j,k}^{2})}{N_{0}} \end{pmatrix}\\ & = \sum_{k=1}^{K}\log\left(1+\frac{\sum \limits_{j=1}^{J}h_{j,k}^{2}p_j \mathbb E(s_{j,k}^{2})}{N_{0}}\right). \end{split} 
 \end{equation}

 \textbf{Lemma 1:}
The ergodic channel capacity of VM-SCMA   is given by
\begin{equation}
\small
\begin{aligned} C=&\max  ~ \mathcal {I}(\mathbf {X};\mathbf {y}|\mathbf {H}) \\=&  K \left [{\mathrm {exp}\left ({\frac {N_{0}}{\bar P}}\right)\mathrm {E_1}\left ({\frac {N_{0}}{\bar{P}}}\right)}\right],\! \end{aligned}
\end{equation}
where  $\mathrm {E_n}(x) = \int_{0}^{ \infty }\frac{e^{-xt}}{t^n} \rm dt $ is the exponential integral function and  $\bar P=  \frac{1}{K}\sum \nolimits_{j=1}^{J} d_{j}^{- \alpha   } p_j  $  is the average  receive  power of a subcarrier. The above bound only  holds when the  average received power at  each subcarrier  is the same,  i.e.,
 \begin{equation} 
 \label{c1}
 \small
  \sum \limits _{j \in   \boldsymbol{\boldsymbol{\phi}}_{\widetilde{\mathbf F}}(1) }p_jd_j^{-\alpha} =   \sum \limits _{j \in   \boldsymbol{\boldsymbol{\phi}}_{\widetilde{\mathbf F}}(2)}p_jd_j^{-\alpha}= , \ldots, = \sum \limits _{j \in   \boldsymbol{\boldsymbol{\phi}}_{\widetilde{\mathbf F}}(K)} p_jd_j^{-\alpha}.
\end{equation}

\textit{Proof:}  
The concavity of the logarithmic function  in (\ref{AMI})    gives rise to
\begin{equation}
\begin{aligned}
\small
\label{ShaC}
    I(\mathbf{S};\mathbf{y}\vert\boldsymbol{H}) & \leq  I_{U}(\mathbf{S};\mathbf{y}\vert\boldsymbol{H})  \\
    & = K\log\left(1+\frac{\sum \limits_{k=1}^{K} \sum \limits_{j=1}^{J} {g}_{j}^2d_{j}^{- \alpha   } p_j \mathbb E(s_{j,k}^{2})}{KN_{0}}\right)
\end{aligned}
\end{equation}
 by applying the Jensen’s inequality. The equality of (\ref{ShaC}) holds if and only if (\ref{c1}) is satisfied.  The probability density function of a  squared channel gain $  g_j^2  $ is   $g_{R}(x) = \frac{1}{2} e^{-x^2/2}, x \geq 0$.  Hence, the ergodic channel capacity of VM-SCMA can be expressed as \cite{CNSRLi}
\begin{equation}
\begin{aligned}
\small
   C &= \int_{0}^{+ \infty} I_{U}(\mathbf{S};\mathbf{y}\vert\boldsymbol{H}) g_{R}(x) dx\\
   & = K  {\mathrm {exp}\left ({\frac {N_{0}}{ \bar P\sum \limits_{k=1}^{K} \mathbb E(s_{j,k}^{2})}}\right)\mathrm {E_1}\left ({\frac {N_{0}}{\bar P\sum \limits_{k=1}^{K} \mathbb E(s_{j,k}^{2})}}\right)}.
          \end{aligned}
\end{equation}
Note that the average power of a codeword  is normalized, i.e., $ \sum \nolimits_{k=1}^{K} \mathbb E(s_{j,k}^{2}) = 1$. Hence, this completes the proof of Lemma 1.

  \textit{2) PEP Analysis of VM-SCMA:} 
  Define  a   vector  for $\hat{\mathbf {S}} = \left [ \hat{\mathbf {s}}_{1},  \hat{\mathbf {s}}_{2}, \ldots, \hat{\mathbf {s}}_{J} \right]$, where $\hat{\mathbf {s}}_{j}   \in  \boldsymbol {\mathcal S}_{j},  j=1,2,\ldots,J$.  Let  $  \text{Pr} \{\mathbf{S} \to \mathbf{\hat{S}}\} $ be  the
 unconditional PEP that $\hat{\mathbf {s}} $  is detected
while the actually transmitted codewords is $\mathbf {s}  $.   Assume that a maximum likelihood detector  is employed at the receiver,  the  upper bound on the asymptotic  average SER  (ASER)   is  given  by   \cite{LPCSMA}
\begin{equation}
\small
\label{Aser}
     P({\text {e}})  \leq  \frac {1}{ \prod  \limits_{j=1}^{J} \vert \boldsymbol {\mathcal S}_{j} \vert}  \sum _{\mathbf {S}}\sum _{\hat {\mathbf {S}}\neq \mathbf {S}}  \text{Pr} \{\mathbf{S} \to \mathbf{\hat{S}}\}.
\end{equation}

A tight approximation  of   $  \text{Pr} \{\mathbf{S} \to \mathbf{\hat{S}}\} $  in Rayleigh fading channels  can be computed as follows \cite{LiusPARSE}:
 \begin{equation} 
 \small
 \label{pep2}
  \text{Pr} \{\mathbf{S} \to \mathbf{\hat{S}}\} 
\simeq \frac {1}{12}\prod _{k=1}^{K} \frac {1}{1+\frac {\Delta_{\mathbf {s}\rightarrow \hat {\mathbf {s}}}(k)}{4N_{0}}}+\frac {1}{4}\prod _{k=1}^{K} \frac {1}{1+\frac { \Delta_{\mathbf {s}\rightarrow \hat {\mathbf {s}}}(k)}{3N_{0}}},
 \end{equation}
 where $\Delta_{\mathbf {s}\rightarrow \hat {\mathbf {s}}}(k)  =  \sum \limits _{j=1}^{J}p_j  d_j^{-\alpha}|s_{j,k}-\hat {s }_{j,k}|^{2}$. We  further  define the set $D_{\mathbf {s}\rightarrow \hat {\mathbf {s}}} \triangleq\left \{{k: \Delta_{\mathbf {s}\rightarrow \hat {\mathbf {s}}}(k) \neq 0, 1\leq k \leq K }\right \}$, and let $G_{\mathbf {s}\rightarrow \hat {\mathbf {s}}}$  be the  cardinality of   $D_{\mathbf {s}\rightarrow \hat {\mathbf {s}}}$.  At  sufficiently  high SNRs,  it  follows  that
   \begin{equation} 
 \small
 \label{pep3}
 \begin{aligned}
 & \text{Pr} \{\mathbf{S} \to \mathbf{\hat{S}}\} \\
&\simeq  N_{0}^{- G_{{{\mathbf S}} \rightarrow {\tilde{\mathbf{S}}}}}  \left ( \frac {4^{G_{{{\mathbf S}} \rightarrow {\tilde{\mathbf{S}}}}}}{12}  + \frac {3^{G_{{{\mathbf S}} \rightarrow {\tilde{\mathbf{S}}}}}}{4} \right) \prod _{k \in D_{\mathbf {s}\rightarrow \hat {\mathbf {s}}}}{ \Delta_{\mathbf {s}\rightarrow \hat {\mathbf {s}}}(k)}.
 \end{aligned}
 \end{equation}
 Note that the minimum value of   $G_{\mathbf {s}\rightarrow \hat {\mathbf {s}}}$ can be $N$ which corresponds
to the pairwise error event that all of the detected
codewords are correct except a codeword of only one layer.  This holds  true  for  both  VM-SCMA  and  conventional SCMA with  the same  modulation  order.  Hence, we  propose  to  approximate    (\ref{Aser})  as
   \begin{equation}
  \label{ASER2}
  \small
  \begin{aligned}
  &P({\text {e}}) \simeq \frac {1}{\prod \limits_{j=1}^{J}\vert \boldsymbol {\mathcal S}_{j} \vert}   \sum _{\mathbf {S}}\sum _{ 
 \hat {\mathbf {S}}\neq \mathbf {S}, G_{{{\mathbf X}} \rightarrow {\tilde{\mathbf{S}}}} = N}   \text{Pr} \{\mathbf{S} \to \mathbf{\hat{S}}\}\\
& = \frac {N_{0}^{- N}}{J}   \frac {4^N \!+2\times 3^N}{12} \times \\
& \quad  \quad   \quad
   \sum _{j=1}^{J}  \frac{d_j^{\alpha  N}}{p_j^N} \underbrace{ \frac{1}{\vert \boldsymbol {\mathcal S}_{j} \vert} \sum _{\mathbf s_{j}}  \sum _{ \hat {\mathbf {s}_j}\neq \mathbf {s}_j }  
    \prod _{k\in  \boldsymbol{\boldsymbol{\varphi}}_{\widetilde{\mathbf F}}(j)  } {|s_{j,k}-\hat {s}_{j,k}|^{-2} }}_{ \text{AIPD}(\boldsymbol{\mathcal S }_j)},
  \end{aligned}
    \end{equation}
 where $\text{AIPD}(\boldsymbol{\mathcal S }_j)$  is referred to as the AIPD of  $\boldsymbol{\mathcal S }_j$.  Clearly,  the term  $\frac{d^{\alpha  N}_j}{p_j^N} \text{AIPD}(\boldsymbol{\mathcal S }_j) $ can be regarded as   the ASER of the $j$th user.  It is note that (\ref{ASER2}) only holds tight at sufficiently high SNRs, however, it is still effective to improve the ASER of the $j$th user by improving  $\frac{d^{\alpha  N}_j}{p_j^N} \text{AIPD}(\boldsymbol{\mathcal S }_j)$, provided that  the  VMM is well designed.

 \textit{Remark 2: (\ref{c1}) is the condition of achieving the asymptotic bound
of sum rate,   derived from the perspective of RNs, i.e, the rows of the factor graph. Different from (\ref{c1}),  (\ref{ASER2}) is derived based on the  BER of VM-SCMA, which can be interpreted by analyzing the PEP  from the perspective of VNs, corresponding to the columns of the factor graph.
 }

 \vspace{-0.3cm}
\subsection{Problem Formulation }
In the proposed VM-SCMA, VM codebooks ($\boldsymbol{\mathcal X}$) are generated according to the VMM.  In addition, once the VM codebooks are obtained, the codebook of the $j$th user  is generated by  $\boldsymbol{\mathcal S }_j    =   \sum \nolimits_{l=1}^{J} w_{j,l}\boldsymbol{\mathcal X}_l , 1 \leq j  \leq J$, where $w_{j,l}$ is the codebook allocation strategy.
 The  objective of VM-SCMA design is to  design the power allocation strategy, VMM, VM codebooks, and the codebook allocation strategy for low error rate performance and large mutual information. It is noted that  the  VMM should be designed subject to a total rate of $R_b$.  This is reasonable since without the rate constraint, the VMM tends to select a lower modulation order since the resultant codebooks generally leads to lower SER performance.  Considering the fairness of different users, we propose to minimize the SER performance  of  the  worst user.  Hence, the  VM-SCMA design  is formulated as  \footnote{ In this paper, statistical channel information     is
considered for VM-SCMA design  instead of instantaneous channel information. This approach
allows the proposed design to benefit from low signaling overhead in VM codebook and
power allocations, making it suitable for practical applications. The VM-SCMA design with   instantaneous instantaneous channel information will be considered in
our future work. }
  \begin{subequations}
 \small
\begin{align}
\mathcal {P}_1: \boldsymbol{\mathcal S}, & {\mathbf p} \; \;       =      \arg\min  \limits_{\boldsymbol{\mathcal S},  {\mathbf p}}  \;     \max \limits_{1\leq  j  \leq  J} \frac{d^{\alpha  N}_j}{p_j^N} \text{AIPD}(\boldsymbol{\mathcal S }_j) {\label{optimization2}}\\
\text {s.t.}  \quad \quad  
& \boldsymbol{\mathcal S }_j    =   \sum \limits_{l=1}^{J} w_{j,l}\boldsymbol{\mathcal X}_l , 1 \leq j  \leq J,\label{Cb}\\
& \sum \limits_{l=1}^{J} w_{l,j} =1, \forall w_{l,j}\in \{0,1\},     \label{wallo}\\
& \text{Tr}\left(\boldsymbol{\mathcal X }_l ^{\mathcal H} \boldsymbol{\mathcal X }_l \right)    =  M_l,  1 \leq l  \leq J, \label{CbP}\\
&  \sum \limits_{j=1}^{J} p_j = J 
 \label{Pt},\\
&  \sum \limits_{l=1}^{J} \log_2 M_l=R_b, R_b \in \mathcal R \label{Rb},\\
 &   \sum \limits _{j \in  \boldsymbol{\boldsymbol{\phi}}_{\widetilde{\mathbf F}}(k')} p_jd_j^{-\alpha}= \sum \limits _{j \in\boldsymbol{\boldsymbol{\phi}}_{\widetilde{\mathbf F}}(k')} p_jd_j^{-\alpha},\label{REP1}  \\
 &   1 \leq k',k \leq K, \quad {d_{\min}} <  \forall {d_{j}}\leq  {d_{\max}}.\label{REP2}
\end{align}
\end{subequations}

 The modulation order  of $\boldsymbol{\mathcal X }_l$ is determined by the VMM, i.e., $\vert \boldsymbol{\mathcal X}_l \vert = M_l$. The codebook allocation strategy is given by (\ref{Cb}) and (\ref{wallo}).  (\ref{CbP})  ensures that the average   power of each   codeword is normalized to unit, and we assume the total transmitted power is $J$, i.e.,   (\ref{Pt}). (\ref{Rb}) denotes that the overall   data rate   is constrained  to  $ R_b \in \mathcal R$, where $\mathcal R$ is  the data rates which can  be supported by VM-SCMA.  (\ref{REP1}) and (\ref{REP2}) are the constraints in (\ref{c1}), which  ensure the VM-SCMA can attain large  mutual information.



\textit{Remark 3:}    \textit{$\mathcal {P}_1$ can be regarded as  a generalized  codebook design problem since the power factors can be merged into  the codebooks.
  It is noted that unlike existing codebook design schemes which consider the MPD as a design metric for Rayleigh fading channels, we mainly employ  the proposed AIPD and the constraint derived by the asymptotic bound of sum rate, i.e., (\ref{c1}).
  Due to different modulation orders are supported within a  VMM, the conventional MPD-based criteria is inapplicable as it is derived for sparse codebooks with the same modulation order.}
  

\vspace{-0.3cm}
\section{ VM-SCMA:  Detailed  Design}
 \label{Design}
In this section, we investigate  the detailed design  of   VM-SCMA.     We first introduce the design of MCs with various modulation orders.  Then,  a generalized VM-SCMA, which can accommodate arbitrary modulation orders,  is proposed by relaxing the     constraints (\ref{REP1}) and  (\ref{REP2}) in $\mathcal {P}_1$.  Finally, optimal  VM-SCMA that satisfies  constraints (\ref{Cb})-(\ref{REP2}) is   introduced.


\begin{figure}[]
	\centering
	\begin{subfigure}{0.16 \textwidth}
  \includegraphics[width=0.95 \linewidth]{{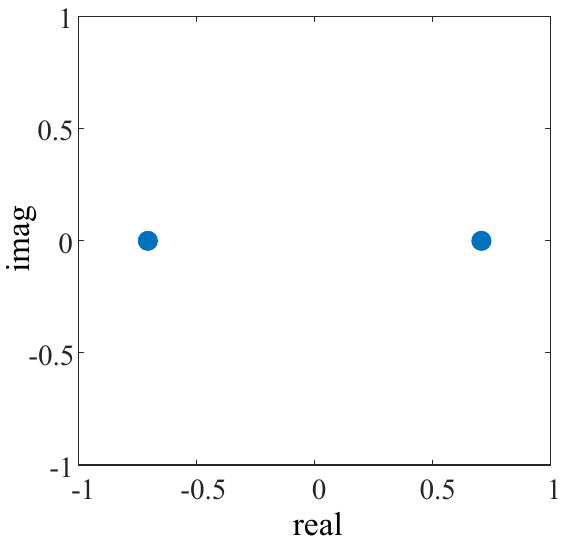}}
		\caption{BPSK  }
				\vspace{-0.1em}
	\end{subfigure}
	\begin{subfigure}{0.16\textwidth}
  \includegraphics[width=0.95 \linewidth]{{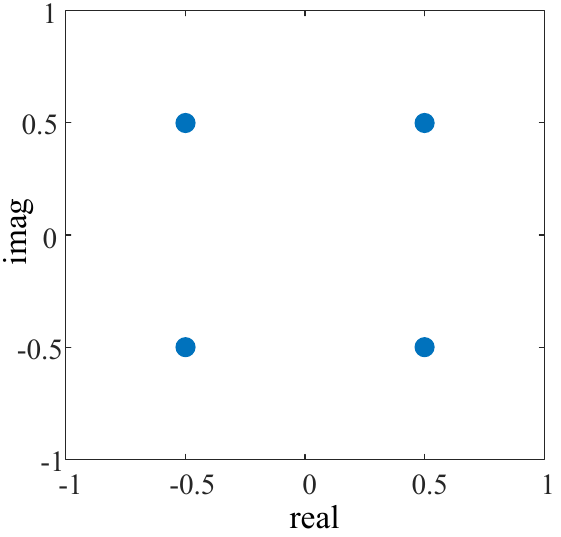}}
		\caption{QPSK  }
		\vspace{-0.1em}
	\end{subfigure}
	\begin{subfigure}{0.16 \textwidth}
  \includegraphics[width=0.95\linewidth]{{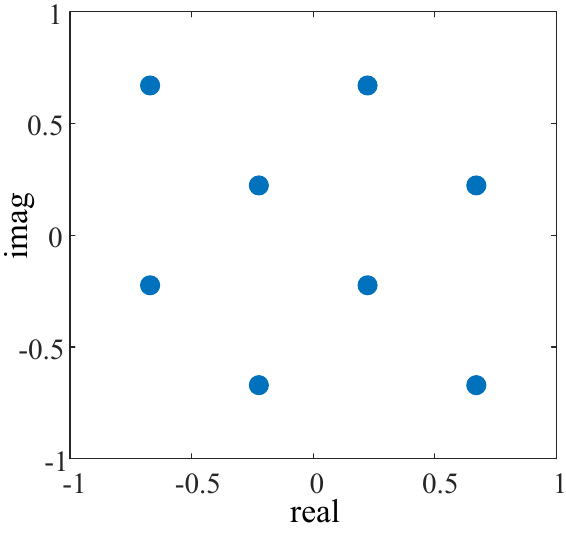}}
		\caption{NS-QAM  }
				\vspace{-0.1em}
	\end{subfigure}
	\begin{subfigure}{0.16\textwidth}
  \includegraphics[width= 0.95 \linewidth]{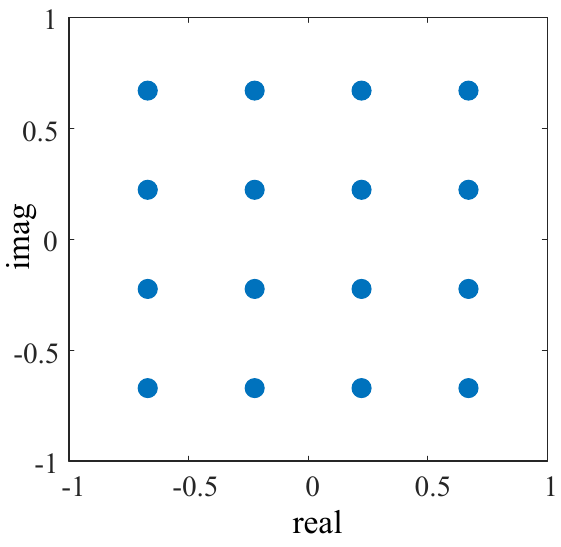}
		\caption{16-QAM}
		\vspace{-0.1em}
	\end{subfigure}	
	\caption{ The employed basic constellation $ \mathbf a $ for $ M = {2, 4, 8, 16}$. }
	\label{basic}
\vspace{-0.3cm}
\end{figure}

 \subsection{ Design  MC pool } 
Let  $ \boldsymbol {\mathcal C}_{\mathcal{MC}}^{M} = \left[\mathbf {{c}}_1^{\mathcal{T}}, \mathbf {{c}}_2^{\mathcal{T}}, \ldots, \mathbf {{c}}_M^{\mathcal{T}} \right]^{\mathcal{T}} \in \mathbb C^{N \times M}$ be an $N$-dimensional MC with its cardinality  given by  $M$. The MCs with different modulation orders constitute an MC pool.  
When considering uplink Rayleigh fading channels, the $l$th  codebook  is generated by 
\begin{equation}
\small
     \boldsymbol{\mathcal X }_l   =    \mathbf V_l     \boldsymbol {\mathcal C}_{\text{MC}}^{M_l},  
\end{equation}
  where     $\mathbf {V}_{l} \in \mathbb {B}^{K \times N} $ is a  mapping   matrix that maps the $N$-dimensional vector  to a $K$-dimensional sparse   codeword. $\mathbf {V}_{l} $    can be constructed  according to the position of the ‘$0$’ elements of  ${{\mathbf{f}}_{l}}$ by inserting  all-zero row vectors into the identity matrix ${{\mathbf{I}}_{N }}$. For example, for the indicator matrix shown in Fig. \ref{factor} ,  we have
\begin{equation} 
\small
\begin{aligned}
{{\mathbf{V}}_{1}}=\left[ \begin{matrix}
  0 & 0  \\
  1 & 0  \\
  0 & 0  \\
  0 & 1  \\
\end{matrix} \right],
{{\mathbf{V}}_{2}}=\left[ \begin{matrix}
  1 & 0  \\
  0 & 0  \\
  0 & 1  \\
  0 & 0  \\
\end{matrix} \right], {{\mathbf{V}}_{3}}=\left[ \begin{matrix}
  1 & 0  \\
  0 & 1  \\
  0 & 0  \\
  0 & 0  \\
\end{matrix} \right],
\\
{{\mathbf{V}}_{4}}=\left[ \begin{matrix}
 0 & 0  \\
  0 & 0  \\
  1 & 0  \\
  0 & 1  \\
\end{matrix} \right],
{{\mathbf{V}}_{5}}=\left[ \begin{matrix}
  1 & 0  \\
  0 & 0  \\
  0 & 1  \\
  0 & 0  \\
\end{matrix} \right],
{{\mathbf{V}}_{6}}=\left[ \begin{matrix}
  0 & 0  \\
  1 & 0  \\
  0 & 1  \\
  0 & 0  \\
\end{matrix} \right].
\end{aligned}
 \end{equation}


  To design  VM-SCMA,    the MC pool needs  to be  determined first.    In this paper, a multi-stage design scheme is proposed to design $\boldsymbol {\mathcal C}_{\text{MC}}$, which mainly includes: a) choose a one dimensional basic constellation, denoted as $ \mathbf a \in \mathbb C^{M \times 1}$, with large MED; b) permute  $ \mathbf a $   to obtain  the $N$-dimensional MC, i.e,   
\begin{equation}
 \small
     \boldsymbol {\mathcal C}_{\text{MC}}^{M}  = \left[ {\pi }_{1} \left( \mathbf a      \right),{\pi }_{2} \left( \mathbf a  \right),\ldots,{\pi }_{N} \left( \mathbf a  \right)\right]^{\mathcal{T}},
\end{equation}
where ${\pi }_{n}$ denotes the permutation operator in the $n$th dimension. The permutation metric is the AIPD of the MC, i.e.,
\begin{equation}
\small
     \Upsilon _{\mathrm{ Per}}=  \frac{1}{ M} \sum _{i=1}^{M}  \sum _{ m =1,  m\neq i}^{M}  \;
    \prod _{n= 1 }^N  |c_{n,i}- {c }_{n,m}|^{-2}, 
\end{equation}
where $c_{n,i} $ denotes the $i$th symbol   the $n$th entry of $  \boldsymbol {\mathcal C}_{\text{MC}}^{M}$.   Considering the decoding complexity of MPA, we only consider   $M= 2,4,8, $ and $16$.  The basic constellations employed are the binary phase-shift keying (BPSK), quadrature PSK (QPSK), non-squared QAM (NS-QAM)  and 16-QAM, as shown in Fig. \ref{basic}.

\subsection{Generalized VM-SCMA}

 After designing the MC pool, it is still challenging to directly solve the problem of $\mathcal {P}_1$ due to the non-convex constraints.  In this subsection, we propose to address $\mathcal {P}_1$ by dividing it into several sub-problems, namely,   power allocation, codebook allocation and VMM design.
 
\textit{1) Power allocation:}   We  start by  investigating the power allocation scheme under the fixed VMM and codebook allocation. Assuming that the codebooks $\boldsymbol{\mathcal S }_j, j=1,2,\ldots,J$ are already determined, the power allocation of the codebook is then formulated as
 \begin{subequations}
  \small
\begin{align}
\mathcal {P}_{1-2}:  {\mathbf p} \; \;       =     & \arg\min  \limits_{ {\mathbf p}}  \;     \max \limits_{1\leq  j  \leq  J} \frac{d^{\alpha  N}_j}{p_j^N} \text{AIPD}(\boldsymbol{\mathcal S }_j) \\
\text {s.t.}  & \quad \quad    (\ref{Pt}),  (\ref{REP1}),  (\ref{REP2}).
\end{align}
\end{subequations}
It it noted that  (\ref{REP1}) and  (\ref{REP2}) are strong constraints involve   power allocation, codebook allocation and VMM design. Here, we first investigate  the power allocation strategy under the  total power constraint, i.e., (\ref{Pt}). The detailed investigation of constraints (\ref{REP1}) and  (\ref{REP2}) will be discussed later. 
Obviously,  the  power allocation strategy can be obtained by solving the following linear equations:
 \begin{equation} 
 \small
 \label{LinEq}
 \begin{aligned}
\Xi =\frac{d^{\alpha  }_1}{p_1} \sqrt[N]{\text{AIPD}(\boldsymbol{\mathcal S}_1 )} = & \frac{d^{\alpha  }_2}{p_2} \sqrt[N]{\text{AIPD}(\boldsymbol{\mathcal S }_2)} \\
 & =, \ldots, = \frac{d^{\alpha  }_J}{p_J} \sqrt[N]{\text{AIPD}(\boldsymbol{\mathcal S}_J )}.
 \end{aligned}
 \end{equation}
  Under the total power constraint  of (\ref{Pt}), we have
  \begin{equation} 
  \small
  \label{P_allo}
p_j = \frac{d_j^{\alpha} \sqrt[N]{\text{AIPD}(\boldsymbol{\mathcal S}_j )}}{\sum \nolimits_{j=1}^{J} d_j^{\alpha} \sqrt[N]{\text{AIPD}(\boldsymbol{\mathcal S }_j)}} J,  \quad j = 1,2,\ldots, J.
  \end{equation}

\textit{2) Codebook allocation:}  Then, we investigate codebook allocation scheme under the fixed  power allocation strategy and VMM. Substituting (\ref{P_allo}) into (\ref{LinEq}),  we have 
\begin{equation}
\small
\label{LinEq2}
\begin{aligned}
      \Xi  &=  \frac{1}{J}\sum \nolimits_{j=1}^{J} d_j^{\alpha} \sqrt[N]{\text{AIPD}(\boldsymbol{\mathcal S }_j)}\\
      &=
  \frac{1}{J} \sum \nolimits_{j=1}^{J} d_j^{\alpha}     \sum \nolimits_{l=1}^{J} w_{j,l}  \sqrt[N]{\text{AIPD}(\boldsymbol{\mathcal X }_l)}. 
\end{aligned}
\end{equation}

As discussed in Section \ref{Fundamentals}, it is effective to improve the error rate performance of  a single user   by minimizing  $\Xi $. Hence,   the   codebook allocation problem  is  formulated as
  \begin{subequations}
\label{optimization3}
 \small
\begin{align}
 \mathcal {P}_{1-3}:  {\mathbf  W }        =     & \arg\min  \limits_{ \mathbf  W}  \quad   \Xi  \tag{\ref{optimization2}}\\
\text {s.t.}  &  \quad   (\ref{wallo}).
\end{align}
\end{subequations}
 Clearly, $ \Xi $ achieves the minimum value when a codebook with larger AIPD value is assigned for the user with smaller $d_j$. Namely,  the user closer to the BS should be assigned a codebook with a higher modulation order, as it enjoys a larger AIPD value.


 \textit{3) VMM design:} Under the power allocation scheme of (\ref{P_allo}),  (\ref{REP1}) and (\ref{REP2}) can be simplified to
 \begin{equation}
 \small
 \begin{aligned}
\label{CLNeq2}
     \sum \limits _{j \in   \boldsymbol{\boldsymbol{\phi}}_{\widetilde{\mathbf F}}(k')} \sqrt[N]{\text{AIPD}(\boldsymbol{\mathcal S }_j)}  = \sum \limits _{j \in  \boldsymbol{\boldsymbol{\phi}}_{\widetilde{\mathbf F}}(k)} \sqrt[N]{\text{AIPD}(\boldsymbol{\mathcal S}_j )},  1 \leq k,k' \leq K.
     \end{aligned}
\end{equation}

    It is noted that the codebook allocation does not affect the  set of codebooks associated with the $k$th RN in a VMM.
Hence,   (\ref{CLNeq2}) can be reformulated as 
 \begin{equation}
\small
\label{CLNeq3}
     \sum \limits _{j \in   \boldsymbol{\boldsymbol{\phi}}_{\mathbf F}(k')} \sqrt[N]{\text{AIPD}(\boldsymbol{\mathcal S }_j)}= \sum \limits _{j \in  \boldsymbol{\boldsymbol{\phi}}_{\mathbf F}(k)} \sqrt[N]{\text{AIPD}(\boldsymbol{\mathcal S }_j)},  1 \leq k,k' \leq K.
\end{equation}

 Since the codebooks  with different modulation orders   have   different   $\text{AIPD} $ values,    (\ref{CLNeq2}) holds only if    the   set of  modulation orders       associated  with the   $k$th RN are the same. Denoted  ${{\mathcal {M}}}(k)=
\left\{ {M_{l}\left| {{f_{k,l}} = 1, l =1,2,\ldots,J} \right.} \right\}$   by the   modulation orders being associated with the $k$th RN in a VMM.  
Namely, the following should hold:
\begin{equation}
\small
\label{MLNeq}
 {{\mathcal {M}}}(1) = {{\mathcal {M}}}(2)= ,\ldots,={{\mathcal {M}}}(K).
\end{equation}
Unfortunately, (\ref{MLNeq}) may not be   achieved for all VMMs.   For example, (\ref{MLNeq}) doesn't hold when the number of different  modulation orders  needs to be assigned to a VMM is larger than $d_f$,      Hence,  the  goal is transformed to minimize the following metric:  
\begin{equation}
\small
\label{CLNeq}
    \tau   =  \max  \limits_{1\leq  k <  k'  \leq K }  \left |      \sum \limits _{l \in   \boldsymbol{\boldsymbol{\phi}}_{\mathbf F}(k')} \sqrt[N]{\text{AIPD}(\boldsymbol{\mathcal X }_l)} -\sum \limits _{l \in   \boldsymbol{\boldsymbol{\phi}}_{\mathbf F}(k)} \sqrt[N]{\text{AIPD}(\boldsymbol{\mathcal X}_l )}  \right|.
\end{equation}

For a given  $\mathfrak m $, the number of     allocations of $\mathfrak m $ to VMM is $J!$. To obtain the optimal allocation, exhaustive search is computationally feasible for small values of $J$ as the VMM can be designed offline.  For large $J$, we propose a  layer switch algorithm  to reduce the   the computational complexity of obtaining VMM.   The proposed detailed design for generalized  VM-SCMA is summarized  in Algorithm \ref{labelingAlg}.
 
 \begin{algorithm}[t] 
\caption{Design generalized VM-SCMA}
\label{labelingAlg}
\begin{algorithmic}[1]
\REQUIRE {$\mathbf F$, $R_b$, MCs pool  and  $\mathbf d$.} \\
\ENSURE {Codebook  $\boldsymbol{\mathcal X}_{j} $ and power allocation   } \\
\STATE{Obtain all the modulation combinations $\mathcal M_c$} that achieve the  required rate $R_b$.
\FOR {for each $\mathfrak m$ in $\mathcal M_c$ }

 \STATE{  \textbf{Step 1: Design VMM}  }  \\
 \STATE{ Randomly allocate the modulation order  in  $\mathfrak m$   to  ${{\mathbf{M}}_{K\times J}}$, and calculate the cost $\tau_{\text{ini}}$ based on  \ref{CLNeq}). Let $\tau =\tau_{\text{ini}}$. }  \\
\FOR {$l=1:J$}  
\FOR {$l'=l:J$}  
\STATE { Switch the $l$th row with $l'$th row  of ${{\mathbf{M}}_{K\times J}}$  and calculate the switched cost $\tau_{\text{sw}}$.   }
\IF{$\tau_{\text{sw}} \le \tau $}
\STATE { Retain the switch, and update  $\tau $  with $\tau_{\text{sw}}$.  }
\ELSE
\STATE { Switch back the two rows.  }
\ENDIF
\ENDFOR
\ENDFOR
 \STATE{  \textbf{Step 2: Generate VM codebooks} }  \\
 \STATE{ Generate the $J$ sparse  codebooks based on the VMM and $\mathbf{\Phi}$, i.e, $ \boldsymbol{\mathcal X }_l   =    \mathbf V_l     \boldsymbol {\mathcal C}_{\text{MC}}^{M_l}, l= 1, 2, \ldots, J$.}
  \STATE{  \textbf{Step 3: Perform codebook allocation} }  \\
  
 \STATE{  Allocate the  $J$ codebooks to   $J$ users under  the rule of the user with small $d$ should  assigned with a codebook with higher modulation order.  }
 
 \STATE{  \textbf{Step 4: Perform power allocation}  }  \\
 \STATE{ Perform the power allocation according to (\ref{P_allo})}.
 \STATE{ Calculate and store $\Xi$} according to (\ref{LinEq2}).
\ENDFOR
 \STATE{Choose the  ${{\mathbf{M}}_{K\times J}}$   that results in the smallest $\Xi$, and output the $J$ codebooks and the power allocation vector. }
 \end{algorithmic}
 \vspace{-0.1cm}
\end{algorithm}


\textit{Example 1:} We now give an example to illustrate  the proposed generalized VM-SCMA. Specifically, we consider  the VM-SCMA with  $R_b = 17$ bits. There are different values of $\boldsymbol {\mathfrak{m}}$  to achieve this rate, such as  $\mathfrak m = \{ 2, 2, 8, 16,16, 16\}$   and $\mathfrak m = \{ 4, 8, 8, 8,8, 8\}$. For  each  $\mathfrak m$, the  VMM can also be different. In particular, we consider the following  four VMMs: 
     \begin{equation} 
           \small    
 \label{Model1}
       \setlength{\arraycolsep}{1.5pt} 
 \begin{aligned}
 &{{\mathbf{M}} _{4\times 6}^{(3)}}=\left[ \begin{matrix}
  0 & 2 & 8 & 0 &  16 & 0  \\
  2 & 0 & 8 & 0 & 0 & 16\\
  0 & 2 & 0 & 16 & 0 & 16 \\
  2 & 0 & 0 & 16 & 16 & 0  \\
\end{matrix} \right],
{{\mathbf{M}}_{4\times 6}^{(4)}}=\left[ \begin{matrix}
  0 & 16 & 16 & 0 &  16 & 0  \\
  2 & 0 & 16 & 0 & 0 & 8\\
  0 & 16  & 0 & 2 & 0 & 8 \\
  2 & 0  & 0 & 2 & 16 & 0  \\
\end{matrix} \right],\\
 &{{\mathbf{M}}_{4\times 6}^{(5)}}=\left[ \begin{matrix}
  0 & 4 & 8 & 0 &  16 & 0  \\
  2 & 0 & 8 & 0 & 0 & 16\\
  0 & 4 & 0 & 8 & 0 & 16 \\
  2 & 0 & 0 & 8 & 16 & 0  \\
\end{matrix} \right],
{{\mathbf{M}}_{4\times 6}^{(6)}}=\left[ \begin{matrix}
  0 & 8 & 8 & 0 &  8 & 0  \\
  4 & 0 & 8 & 0 & 0 & 8\\
  0 & 8 & 0 & 8 & 0 & 8 \\
  4 & 0 & 0 & 8 & 8 & 0  \\
\end{matrix} \right]. 
 \end{aligned}
       \end{equation} 
 ${{\mathbf{M}}_{4\times 6}^{(3)}}$ and   ${{\mathbf{M}}_{4\times 6}^{(4)}}$ have the same   modulation order combination, i.e., $\mathfrak m = \{2, 2, 8, 16,16, 16 \}$. The set of modulation orders associated with the first RN in  ${{\mathbf{M}}_{4\times 6}^{(3)}}$ and   ${{\mathbf{M}}_{4\times 6}^{(4)}}$  are given by ${{\mathcal {M}}}(1) = \{ 2,8,16\}$ and ${{\mathcal {M}}}'(1) = \{ 16,16,16\}$, respectively. Clearly, according to  (\ref{CLNeq}), 
   ${{\mathbf{M}}_{4\times 6}^{(3)}}$  is better than that of ${{\mathbf{M}}_{4\times 6}^{(4)}}$  as it  own smaller value of $  \tau $.     ${{\mathbf{M}}_{4\times 6}^{(5)}}$ and   ${{\mathbf{M}}_{4\times 6}^{(6)}}$ are the optimized VMMs for  $\mathfrak m =  \{2, 4, 8, 8,16, 16\}$ and $\mathfrak m = \{ 4, 8, 8, 8,8, 8\}$, respectively.   
  For a given $R_b$, the optimum VMM may  vary with  $\mathbf d$.     
 Later on, we will show   ${{\mathbf{M}}_{4\times 6}^{(6)}}$ is  better than ${{\mathbf{M}}_{4\times 6}^{(3)}}$ and ${{\mathbf{M}}_{4\times 6}^{(5)}}$ in the case of $ d_i = d_j, 1\leq i,j\leq J$. However, for $ \mathbf d = [4.7,4.6, 1.6,1.2, 1.1 ]$,  ${{\mathbf{M}}_{4\times 6}^{(5)}}$   achieves the best error rate performance.      
  
  Upon determining the optimum VMM for a given rate $R_b$, the sparse codebook is obtained by $   \boldsymbol{\mathcal X }_l   =    \mathbf V_l     \boldsymbol {\mathcal C}_{\text{MC}}^{M_l}, l=1,2,\ldots,J$. Then,  a codebook with higher modulation order is allocated to a user with a smaller distance.   For ${{\mathbf{M}}_{4\times 6}^{(3)}}$,   ${{\mathbf{M}}_{4\times 6}^{(5)}}$ and   ${{\mathbf{M}}_{4\times 6}^{(6)}}$, the   codebook allocation is given by  ${{\mathbf{W}}} = \mathbf{I}_{J}$.  Finally, the power allocation strategy in (\ref{P_allo}) is applied to $J$ codebooks.


\textit{Remark 4:} \textit{To sum up,  the proposed generalized VM-SCMA (Algorithm 1) has the following features:
\begin{itemize}
    \item      The proposed generalized VM-SCMA can be implemented with any arbitrary indicator matrix. 
           \item The proposed generalized VM-SCMA enables a wide range of rates, i.e,   $ R_b \in \{6,7,\ldots, 24\}$. For a given $R_b $, the   sparse codebook  that achieves the beset error rate performance can   be determined according  Algorithm 1.
       \item Algorithm 1 can also be employed to design the codebook with arbitrary  modulation order combinations.    
\end{itemize}
 }

\vspace{-0.3cm}
\subsection{Optimal VM-SCMA}
  We now consider a special case, where (\ref{MLNeq}) is satisfied,  i,e,  $\tau  =0$. Such VM-SCMA is referred to as optimal VM-SCMA.

\textbf{Lemma 2}: For an indicator matrix $\mathbf F_{K \times J}$, if the $J$ columns can be divided into $d_f$ groups with the $i$th group denoted by $G_i$,      and $\sum  \nolimits_{l \in G_i} \mathbf  f_l = \mathbf 1^{K \times 1}, 1\leq i \leq d_f$,  then  $\mathbf F_{K \times J}$ can be designed to achieve  optimal VM-SCMA and up to $d_f$  different modulations  can be accommodated.

\textit{Proof:} Note that $\sum  \nolimits_{l \in G_i} \mathbf  f_l = \mathbf 1^{K \times 1}, 1\leq i \leq d_f$. By assigning the same modulation order to $i$th group, the resultant  VMMs can achieve $\tau  =0$. 

Consider the  $\mathbf F_{4 \times 6}$ presented  in Fig. \ref{factor}, the columns can be divided into $d_f=3$ groups, i.e.,
\begin{equation}
\small
   \mathbf F_{4 \times 6} =  \Biggg  [ 
   \rule{0cm}{1cm}
   \underbrace{
\begin{array}{cc:c}
  0 & 1\\
  1 & 0 \\
  0 & 1 \\
  1 & 0  
      \end{array}}_{G_1} 
        \underbrace{      \begin{array}{cc:c}
     1 & 0     \\
    1 & 0 \\
     0 & 1   \\
     0 & 1     
      \end{array}}_{G_2} 
                   \underbrace{\begin{array}{ccc}
   1 & 0  \\
      0 & 1\\
       0  & 1 \\
       1 & 0  
      \end{array}}_{G_3} 
       \Biggg   ].
\end{equation}

The   optimal VM-SCMA can  also be extended to  the  SCMA system with larger size indicator matrix. For example,  considering $K=6,J=9,N=2,d_f=3$, the grouped  indicator matrix is given by 
\begin{equation}
\small
   \mathbf F_{6 \times 9} =  \Bigggg [ 
   \rule{0cm}{1cm}
   \underbrace{
\begin{array}{ccc:}
  0 & 0 & 1   \\
  0 & 1 & 0  \\
  0 & 0 & 1  \\
  1 & 0 & 0 \\
  1 & 0 & 0  \\
  0 & 1 & 0  \\
      \end{array}}_{G_1} 
        \underbrace{      \begin{array}{ccc:}
  0 & 1 & 0   \\
  1 & 0 & 0  \\
  1 & 0 & 0  \\
  0 & 1 & 0 \\
  0 & 0 & 1  \\
  0 & 0 & 1  
      \end{array}}_{G_2} 
                   \underbrace{\begin{array}{ccc}
  1 & 0 & 0   \\
  1 & 0 & 0  \\
  0 & 0 & 1  \\
  0 & 1 & 0 \\
  0 & 0 & 1  \\
  0 & 1 & 0  
      \end{array}}_{G_3} 
       \Bigggg  ].
\end{equation}

\textit{Example 2:} We now given an example to illustrate the designed  optimal VM-SCMA. Specifically, we consider the following  optimum VMMs:
        \begin{equation} 
           \small    
 \label{ModelNE}
       \setlength{\arraycolsep}{1.6pt} 
 \begin{aligned}
 &{{\mathbf{M}}_{4\times 6}^{(7)}}=\left[ \begin{matrix}
  0 & 8 & 8 & 0 &  8 & 0  \\
  8 & 0 & 8 & 0 & 0 & 8\\
  0 & 8 & 0 & 8 & 0 & 8 \\
  8 & 0 & 0 & 8 & 8 & 0  \\
\end{matrix} \right],
{{\mathbf{M}}_{4\times 6}^{(8)}}=\left[ \begin{matrix}
  0 & 4 & 8 & 0 &  16 & 0  \\
  4 & 0 & 8 & 0 & 0 & 16\\
  0 & 4 & 0 & 8 & 0 & 16 \\
  4 & 0 & 0 & 8 & 16 & 0  \\
\end{matrix} \right],\\
 &{{\mathbf{M}}_{4\times 6}^{(9)}}=\left[ \begin{matrix}
  0 & 2 & 16 & 0 &  16 & 0  \\
  2 & 0 & 16& 0 & 0 & 16\\
  0 & 2 & 0 & 16 & 0 & 16 \\
  2 & 0 & 0 & 16& 16 & 0  \\
\end{matrix} \right],
{{\mathbf{M}}_{4\times 6}^{(10)}}=\left[ \begin{matrix}
  0 & 8 & 8 & 0 & 16 & 0  \\
  8 & 0 & 8 & 0 & 0 & 16\\
  0 & 8 & 0 & 8 & 0 & 16 \\
  8 & 0 & 0 & 8 & 16 & 0  \\
\end{matrix} \right]. 
 \end{aligned}
       \end{equation} 
  As can be seen, $ {{\mathbf{M}}_{4\times 6}^{(7)}},  {{\mathbf{M}}_{4\times 6}^{(8)}} $ and $  {{\mathbf{M}}_{4\times 6}^{(9)}}$ have the same rate  but different $\mathfrak m $.   The set of modulation orders   are the same at each RN in $ {{\mathbf{M}}_{4\times 6}^{(7)}},  {{\mathbf{M}}_{4\times 6}^{(8)}} $ $  {{\mathbf{M}}_{4\times 6}^{(9)}}$ and $  {{\mathbf{M}}_{4\times 6}^{(10)}}$, leading to $\tau =0$.  Note  that ${{\mathbf{M}}_{4\times 6}^{(7)}}$ with  the same modulation order is generally considered in the literature, which is  a  special case of the proposed  optimum VM-SCMA. The  overall rate for  $  {{\mathbf{M}}_{4\times 6}^{(10)}}$ is given  by $R_b =20$ bits,  which cannot be supported by the  existing sparse codebooks.



\section{The proposed AVM-SCMA}
\label{AVMSec}

In Section \ref{VM-SCMA}, the proposed VM-SCMA  not only   supports diverse overall date rates but also enables variable modulation orders within an SCMA group. Naturally, a fundamental problem is how to adaptively  select the VMM according  to   the users'  statistical SNRs. We will solve this problem    by maximizing the effective throughput  of the VM-SCMA  systems subject to a   reliable constraint. 

\subsection{Effective Throughput of VM-SCMA}
To evaluate the SCMA system performance, most of the
previous works adopt metrics based on the Shannon capacity with Gaussian inputs \cite{CheraghyJoint,Jaber1,Evangelista}. However, the transmit signals in practice are generally constrained to discrete  codebooks. As such, the resulting schemes may not achieve the optimum performance. To tackle this issue, we employ the effective throughput as the performance metric.  The effective symbol rate of the $j$th user, which  measures the correctly
transmitted symbol rate, can be expressed as $1-\text{SER}_{j}$.  Accordingly, the effective throughput of the $j$th user  is defined as $(1-\text{SER}_{j})\log _{2}\vert \boldsymbol{\mathcal S}_j \vert$  \cite{EffectiveWang}. Then, the overall effective throughput for the VM-SCMA system can be expressed as 
\begin{equation} 
\small
\label{SEOpt}
T_{\text{SCMA}}\triangleq \sum \limits_{j=1}^{J} (1-\text{SER}_{j})\log _{2}\vert \boldsymbol{\mathcal S }_j\vert. \end{equation}

Hence, the   adaptive modulation problem for VM-SCMA  is formulated as:    
     \begin{subequations}
 \small
 \label{opt3}
 \begin{align}
       \boldsymbol{\mathcal S},  {\mathbf p} \; \; \;    =     & \arg\max\limits_{\boldsymbol{\mathcal S},  {\mathbf p}}  \;\; \;  T_{\text{SCMA}},  \tag{\ref{opt3}}\\
       \text {s.t.}   \; \; \;  &    \begin{array}{l}
      \text{SER}_{j} \leq  \text{SER}_\text{th}, j=1,2,\ldots,J,
     \end{array} 
     \end{align}
 \end{subequations}
where $\text{SER}_\text{th}$ is the required SER threshold. 


  \begin{table}[]
  \small
 \centering
 \caption{VM codebooks employed in AVM-SCMA. }
     \begin{tabular}{c|c|c|c}
    \hline
    \hline
       \makecell[c] { $R_b$}    &  \text{TM} ($v$)   &  $\mathbf m$ &  $ \gamma_{v}^{\text{th}},a_{v}, b_{v}$  \\
    \hline
     $ 6$  &$\text{TM}1$ & $\left [ 2,2,2,2,2,2\right]$ & $4, 0.42,0.83 $ \\
     \hline
     $8$  & $\text{TM}{2}$ & $\left [ 2,2,2,2,4,4\right]$ & $6, 0.44, 2.57 $ \\
     \hline
         \multirow{2}{*}{$ 10$  }
          & $\text{TM}{3}$ & $\left [ 2,2,4,4,4,4\right]$ &  $8, 0.45, 6.79 $ \\
           \cline{2-4}
           & $\text{TM}{4}$ & $\left [ 2,2,2,2,8,8\right]$ &  $8, 0.43, 6.79 $ \\
     \hline
           \multirow{3}{*}{$ 12$  }
          & $\text{TM}{5}$ & $\left [ 4,4,4,4,4,4\right]$ & $ 10, 0.46, 18.6$ \\
           \cline{2-4}
           & $\text{TM}{6}$ & $\left [ 2,2,4,4,8,8\right]$ & $ 10,0.46, 19.1$ \\
          \cline{2-4}
           & $\text{TM}{7}$ & $\left [ 2,2,2,2,16,16\right]$ & $ 10,0.46, 21.2$ \\
    \hline          
               \multirow{3}{*}{$ 14$  }
          & $\text{TM}{8}$ & $\left [ 4,4,4,4,8,8\right]$ & $ 12, 0.50,72.5$ \\
           \cline{2-4}
           & $\text{TM}{9}$ & $\left [ 2,2,4,4,16,16\right]$ & $ 12,0.49, 68$ \\
          \cline{2-4}
           & $\text{TM}{10}$ & $\left [ 2,2,8,8,8,8\right]$ & $ 12,0.47,61$ \\       
     \hline 
         \multirow{3}{*}{$ 16$  }
          & $\text{TM}{11}$ & $\left [ 4,4,8,8,8,8\right]$ & $ 14,0.52,239$ \\
           \cline{2-4}
           & $\text{TM}{12}$ & $\left [ 4,4,4,4,16,16\right]$ & $14,0.50,195$ \\
          \cline{2-4}
           & $\text{TM}{13}$ & $\left [ 2,2,8,8,16,16\right]$ & $14,0.50,234$ \\       
     \hline    
            \multirow{3}{*}{$18$  }
          & $\text{TM}{14}$ & $\left [ 8,8,8,8,8,8,8\right]$ & $ 16,0.57,1515$ \\
           \cline{2-4}
           & $\text{TM}{15}$ & $\left [ 4,4,8,8,16,16\right]$ & $16,0.55,1010$ \\
          \cline{2-4}
           & $\text{TM}{16}$ & $\left [ 2,2,16,16,16,16\right]$ & $ 16,0.52,653$ \\       
     \hline      
                \multirow{2}{*}{ $20$  }
          & $\text{TM}{17}$ & $\left [ 8,8,8,8,16,16\right]$ & $ 18,0.60,8590$ \\
           \cline{2-4}
           & $\text{TM}{18}$ & $\left [ 4,4,16,16,16,16\right]$ & $18,0.58,5253$ \\
     \hline      
$ 22$   & $\text{TM}{19}$ & $\left [ 8,8,16,16,16,16\right]$ & $20,0.65,7369$ \\
     \hline      
$24$   & $\text{TM}{20}$ & $\left [16,16,16,16,16,16\right]$ & $22,0.68,6367$ \\
     \hline      
     \end{tabular}
     \label{VMMM}
 \end{table}

\subsection{The Proposed AVM-SCMA}
\label{adVM}
In  SCMA systems, due to   muti-user interference, the single-user SER is inapplicable to approximate the  true error rate performance, especially in the low-to-mid SNR regime. A tight approximation of  the $j$th user's SER  is given by
\begin{equation}
\small
\label{SERj}
 \text{SER}_{j}  \simeq \frac {1}{ \prod_{j=1}^{J} \vert \boldsymbol{\mathcal S }_j\vert} \sum _{\mathbf {S}} \sum _{\hat{\mathbf {S}}, \mathbf {s}_j \ne \hat{\mathbf {s}}_j }\text{Pr} \lbrace \mathbf {S} \rightarrow \hat{\mathbf {S}} \rbrace, \end{equation}
where $\text{Pr} \lbrace \mathbf {S} \rightarrow \hat{\mathbf {S}} \rbrace$ is given in (\ref{pep2}).
Obviously, the SER performance of the $j$th user is mutually coupled with other users. Hence, the sparse codebooks  should be jointly designed based on the $J$ users' channel conditions in order to maximize the  effective throughput. Most of existing works on multi-user systems with adaptive modulation designs are based on the individual user's SNR thresholds, which are the sub-optimal solutions. To obtain the   optimum solution, the BS has to exhaustively search all the VMMs in order to maximize the $T_{\text{SCMA}}$ in (\ref{SEOpt}). Namely, the BS needs to perform the following steps: 
\begin{enumerate}
    \item  For each  VMM, conduct the proposed power  and codebook allocation, and generate the  codebooks.
       \item  Calculate the  the SER performance, i.e.,  $ \text{SER}_{j}, j=1,2,\ldots,J$  and the $T_{\text{SCMA}}$. 
 \item Choose the VMM that with maximum  $T_{\text{SCMA}}$ under the constraints of  $\text{SER}_{j} \leq  \text{SER}_\text{th}, j=1,2,\ldots,J.$
\end{enumerate}

However,  there  is a  huge number of possible VMMs, and 
the computational complexity of calculating (\ref{SERj}) for each VMM is also prohibitively high. Specifically, for each round of search, it requires approximately $\prod_{l=1}^{J} M_l^2$ times calculation of $ \text {Pr}(\mathbf {S}\rightarrow \hat {\mathbf {S}})  $ to estimate $ \text{SER}_{j}$, which is unaffordable for the BS.  In this paper, we consider a sub-optimal approach   to design the adaptive modulation scheme that can avoid the  exhaustive search and computation of (\ref{SERj}).

 Similar  to existing AM schemes \cite{ EffectiveWang, QingwenCross,Zhendong}, an adaptive transmission table is designed for AVM-SCMA, which is presented in Table \ref{VMMM}.   It is noted that only the  optimal VM codebooks are employed for transmission model (TM)  design,  leading to $V_{\text{max}} = 20$ TMs.  For simplicity, the modulation order in a VMM is represented by a VM vector, denoted by $\mathbf m = [M_1,M_2,\ldots,M_J]$.   

To proceed, we further define the received statistical  SNRs as follows:
\begin{equation}
\small
\label{asnr}
    \gamma =  \frac{\sum \nolimits_{j=1}^{J} p_j \mathbb E \left\{g_j^2\right \} }{N_0} = \frac{J  \sum \nolimits_{j=1}^{J}    \sqrt[N]{\text{AIPD}(\boldsymbol{\mathcal S }_j)}}{N_0 \sum \nolimits_{j=1}^{J} d_j^{\alpha} \sqrt[N]{\text{AIPD}(\boldsymbol{\mathcal S }_j)}}.   
\end{equation}

Then, we introduce the following lemma.

\textbf{Lemma 3:} Consider two user groups with each   consisting  of $J$ users, and the distances are denoted by $\mathbf d_1$ and $\mathbf d_2$, respectively, where $\mathbf d_1 \neq \mathbf d_2$. For  a given VMM, under the proposed codebook and power allocations, the  two user groups achieve the same SER performance if the received    statistical SNRs    are the same.

\textit{Proof:} Under the  power allocation  of  (\ref{P_allo}), the received power for any of two  users in a group satisfies
\begin{equation}
\small
\label{snrj}
   \frac{ p_j \mathbb E \left\{g_j \right \} }{ p_{j'} \mathbb E \left\{p_{j'}\right \}} = \frac{ \sqrt[N]{\text{AIPD}(\boldsymbol{\mathcal S}_j })}{\sqrt[N]{\text{AIPD}(\boldsymbol{\mathcal S}_{j'} )}}, 1 \leq j, j'\leq J,
\end{equation}
which  only depends on the codebooks of the two users. Hence, if  $  \gamma  $ is the same for the two user groups, the received power  of the $j$th user in each group will also maintain  the same under the same codebook  allocation strategy, leading to  the same SER performance. 


        \begin{algorithm}[t] 
\caption{Proposed  AVM-SCMA}
\label{Adapt}
\begin{algorithmic}[1]
\REQUIRE {$\mathbf d$, $N_0$  and Table \ref{VMMM}.} \\
\STATE{Calculate $\gamma'=  \frac{J^2}{N_0^2  \sum \nolimits_{j=1}^{J} d_j^{\alpha}    } $  by assuming  the same modulation order.}
\STATE{Compare $\gamma'$ with $ \gamma_{\text{th}}^1,  \gamma_{\text{th}}^5,  \gamma_{\text{th}}^{14}$ and $\gamma_{\text{th}}^{20}$. The  nearest  TM   is denoted by   $v_{\text{ini}} \in \{1,5,14,20\}$. }
\FOR { $v = v_{\text{ini}}+1 : V_{\text{max}}$ }
 \STATE{For $\text{TM}{v}$, calculate $\gamma =   \frac{J  \sum \nolimits_{j=1}^{J}    \sqrt[N]{\text{AIPD}(\boldsymbol{\mathcal X_j })}}{N_0^2  \sum \nolimits_{j=1}^{J} d_j^{\alpha} \sqrt[N]{\text{AIPD}(\boldsymbol{\mathcal X_j })}}$.}
  \STATE{Obtain $\text{SER}_{j}^{v} $ based on $\gamma $ and compare $\text{SER}_{j}^{v} $ with  $\text{SER}_\text{th}$.  }
    \STATE{ Calculate $T_{\text{SCMA}}$.  }
  \ENDFOR 
   \STATE{Choose the TM that maximize  $T_{\text{SCMA}}$ and $\text{SER}_{j} \leq  \text{SER}_\text{th}$.  }
 \end{algorithmic}
\end{algorithm}

By employing Lemma $3$, we can determine the SER performance for each TM  according to the received statistical SNRs for arbitrary  $\mathbf d$. Specifically,  we consider $\mathbf d_{0} = \mathbf 1^{J \times 1}$ as the  reference distance for each TM.      Similar to the existing AM designs, we  approximate the $\text{SER} $ of the $v$th TM and the $j$th user for $\mathbf d_{0}$ as \cite{Zhendong,SeongDegrees}
\begin{equation}
\small
\label{Ser_APP}
     \text{SER}_{j}^{v} \approx a_{v} e^{-b_{v} \gamma}, \gamma \geq  \gamma^{\text{th}}_v , j =1,2,\ldots, J,
\end{equation}
where $a_{v}, b_{v}$ are  constants and $ \gamma^{\text{th}}_v $ is the SNR threshold. 
The   parameters $a_{v}, b_{v}$   and $ \gamma^{\text{th}}_v $  are obtained by approximating (\ref{pep2}), which are   also presented in Table \ref{VMMM}. Note that  (\ref{Ser_APP}) can be employed to determine   
the  SER performance for arbitrary $\gamma$.
For the given SER threshold, the required received SNR of $v$th TM can be obtained by 
\begin{equation}
\small
\label{SNRTH}
    \gamma^{\text{th}}_v \approx b_{v}^{-1}\ln { \left( a_{v}  \text{SER}_{\text{th}}^{-1} \right)}.
\end{equation}

Based on (\ref{Ser_APP}),  (\ref{SNRTH}) and Lemma $3$, we propose a low complexity  AVM-SCMA scheme   in Algorithm \ref{Adapt}.   
The   SNR  thresholds  of  the TMs  with  the same modulation order, i.e.,   $ \gamma_{\text{th}}^1,  \gamma_{\text{th}}^5,  \gamma_{\text{th}}^{14}$ and $\gamma_{\text{th}}^{20}$,  are    served  as the reference SNR  thresholds to determine the starting point  of the  search.  Specifically,  by assuming   the same  modulation  order,  the  received SNR  is  give by   $\gamma'=  \frac{J^2}{N_0^2  \sum \nolimits_{j=1}^{J} d_j^{\alpha}    } $   according to      (\ref{asnr}). Then, by comparing $\gamma'$ with $ \gamma_{\text{th}}^1,  \gamma_{\text{th}}^5,  \gamma_{\text{th}}^{14}$ and $\gamma_{\text{th}}^{20}$,  the  closest  TM model is selected as the initial search TM. For example, assuming  $  \gamma' \in [\gamma_{\text{th}}^5,  \gamma_{\text{th}}^{14})$,   the BS will search   from the  $ \text{TM}{5}$ to higher TM.   For each TM, the BS first calculate the  received SNR based on (\ref{asnr}), and  then obtain the $ \text{SER}_{j}^{v}$ and $T_{\text{SCMA}}$ based on the proposed SER model. Finally, the BS selects the TM that maximize the $T_{\text{SCMA}}$ and satisfies the SER constraint.

\textit{Remark $5$:}
  \textit{(\ref{SNRTH}) and Lemma $3$ can be further employed to predict the SER performance gain between any two TMs. Specifically,  to achieve the same SER performance of $\text{SER}_{\text{th}}$  for  $\text{TM}{v_1}$  and $\text{TM}{v_2}$, $ 1 \leq v_1, v_2 \leq V_{\text{max}}$, the power difference in dB  can be approximated as 
\begin{equation}
\small
\label{Pgain}
   G \approx  10\log_{10}{\frac{\gamma_{v_1} -  \gamma_{\text{th}}^{v_1} }{\gamma_{v_2} -  \gamma_{\text{th}}^{v_2}}},  
\end{equation}
where $\gamma_{v_1}$ and $\gamma_{v_2}$ are the received SNRs for  $\text{TM}{v_1}$  and $\text{TM}{v_2}$ in (\ref{asnr}), respectively.}


\vspace{-0.2cm}
\section{Numerical Results }
\label{Simr}
In this section, we conduct extensive simulations to evaluate the proposed VM-SCMA. Specifically, Subsection \ref{eqd} presents the SER performance with  the equal distance case, i.e., $\mathbf d =  \mathbf 1^{J \times 1}$, followed by the performance evaluation over diverse distance case in Subsection \ref{dvd}. Finally, the effective throughput performance of the proposed AVM-SCMA scheme is presented in Subsection \ref{etr}.

\vspace{-0.2cm}
 \subsection{Computational Complexity Analysis}

For the proposed VM-SCMA, the   computational complexity is primarily associated with the VM codebook design, as well as the codebook and power allocations. In VM codebook design, the permutation of MC at each dimension has a computational complexity of $\mathcal O(M!)$, which is becomes substantial  for $M\geq 8$. Fortunately, the modified  binary switching algorithm (BSA) proposed in \cite{LPCSMA}  can be employed to reduce the complexity to $\mathcal O(I_{\text{iter}}M^2)$, where ${I_{\text{iter}}}$ denotes the number of iterations in the modified BSA. In addition, the  computational complexity of VMM design in   \textbf{Algorithm  \ref{labelingAlg}}, i.e., Step $1$ can be approximated as $\mathcal O(M^2)$. It is noted that the VM codebook design is an offline   process, and once the VM codebooks are designed, it can be directly deployed for online.   Moreover, the codebook and power allocation has less computational complexity, which are negligible. Since the statistical channel information is employed for VM-SCMA design instead the instantaneous  channel information, the signaling overhead of VM codebook and power allocations are relatively small.

As discussed in Section \ref{AVMSec}, the proposed AVM-SCMA avoids the need to calculate the PEP of (\ref{SERj}) for each VMM and eliminates the exhaustive search of all possible VMMs. Moreover, the search space for the proposed AVM-SCMA in \textbf{Algorithm \ref{Adapt}} is upper bounded by the number of TMs listed in Table \ref{VMMM}. In a nutshell, both the VM-SCMA and AVM-SCMA schemes are computationally feasible for practical applications.

 \label{eqd}
 \begin{figure}[!t]
     \centering
   \includegraphics[width= 0.9 \linewidth]{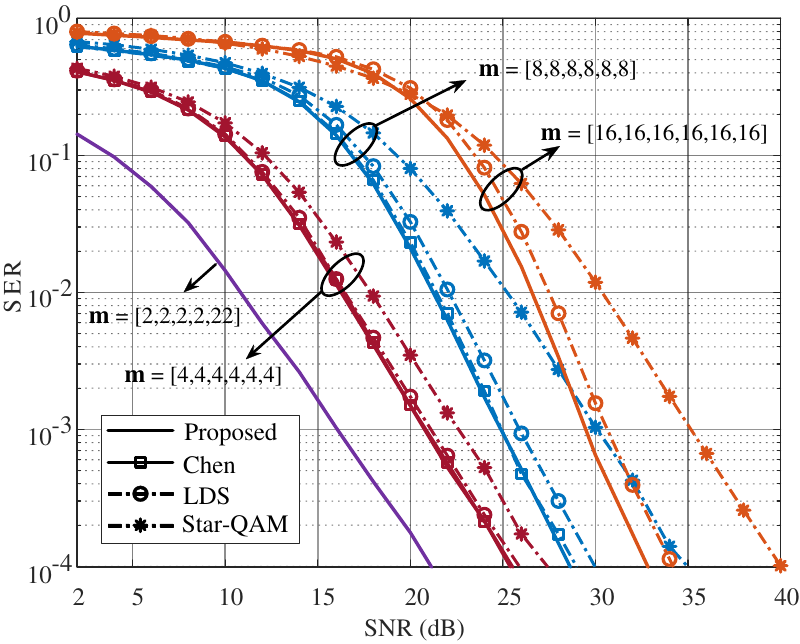}
  \caption{SER performance with  the same modulation order and equal distance.}\label{SER_Meq}
  \vspace{-0.4cm}
 \end{figure}

\begin{figure*}[htbp]
	\centering
	\begin{subfigure}{0.29 \textwidth}
  \includegraphics[width=0.94 \linewidth]{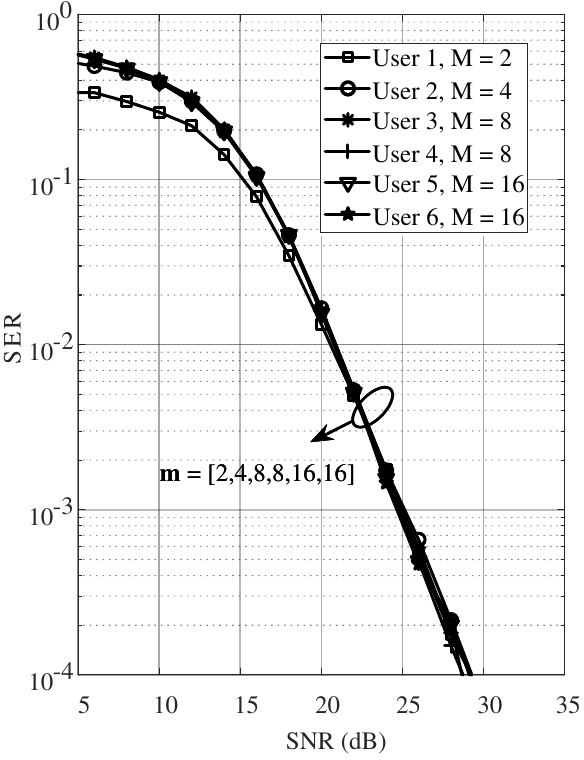}
		\caption{Single user SER ($R_b = 17$ bits).  }
				\vspace{-0.1em}
	\end{subfigure}
	\begin{subfigure}{0.29\textwidth}
  \includegraphics[width=0.95  \linewidth]{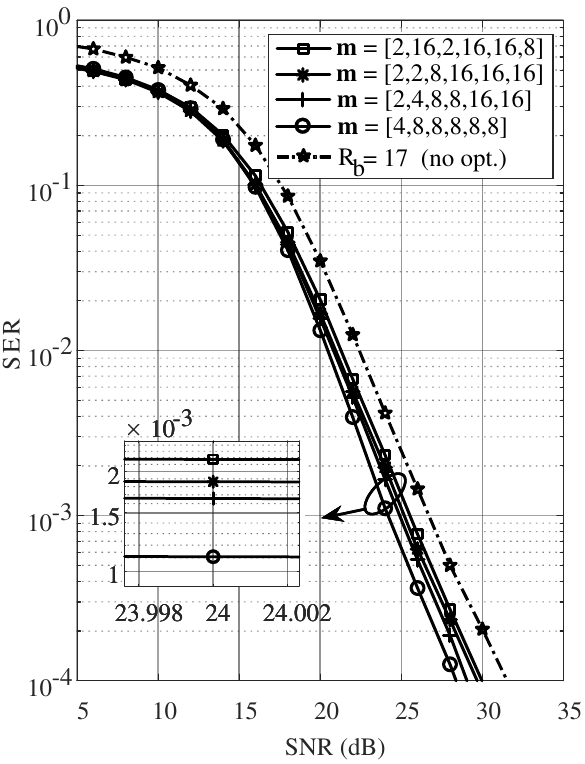}
		\caption{$R_b = 17$ bits.  }
		\vspace{-0.1em}
	\end{subfigure}
	\begin{subfigure}{0.3\textwidth}
  \includegraphics[width=0.93  \linewidth]{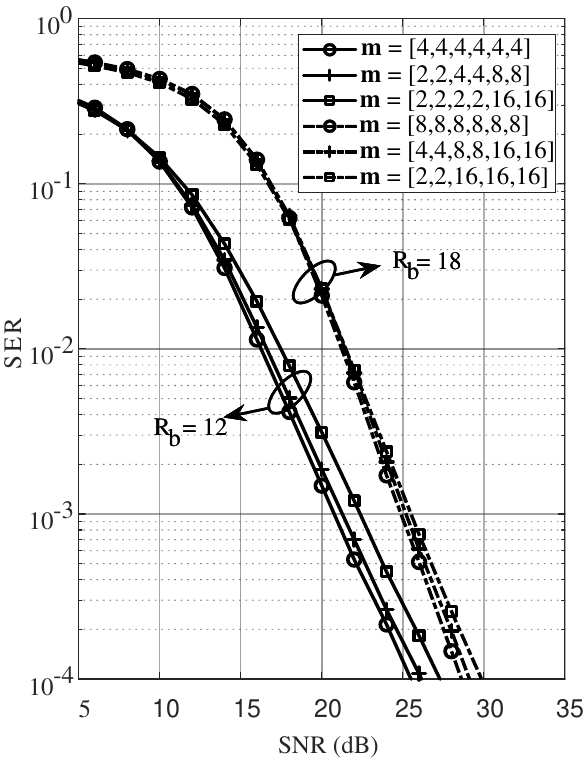}
		\caption{$R_b = 12 $ bits and $18$ bits. }
		\vspace{-0.1em}
	\end{subfigure}	
	\caption{SER performance with   the same distance.}
	\label{SER_eq}
\end{figure*}

\begin{figure*}[htbp]
	\centering
	\begin{subfigure}{0.45 \textwidth}
  \includegraphics[width=1 \textwidth]{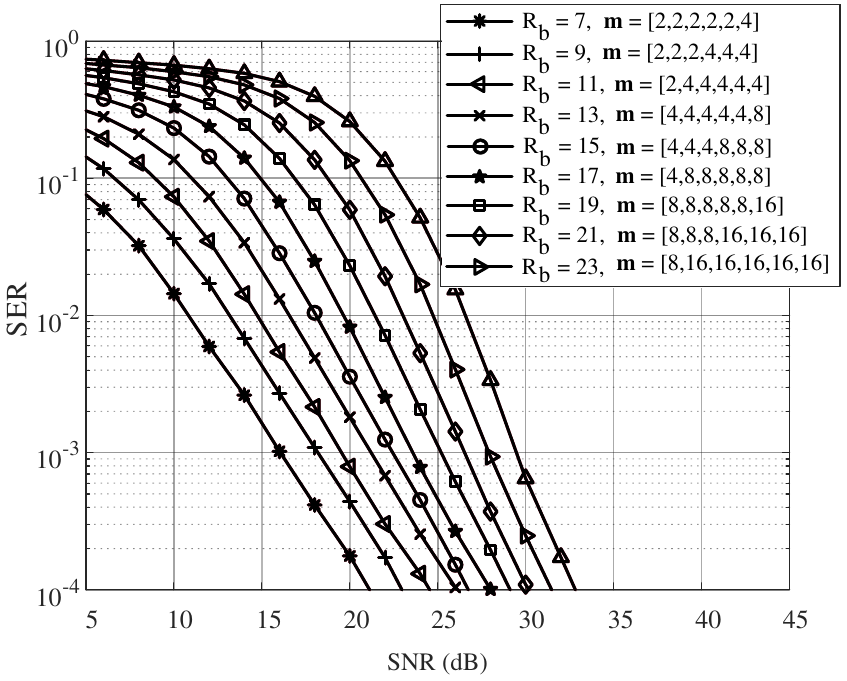}
		\caption{$R_b = 7,9,\ldots, $ and $23 $  }
	\end{subfigure}
	\begin{subfigure}{0.46\textwidth}
  \includegraphics[width= 1 \textwidth]{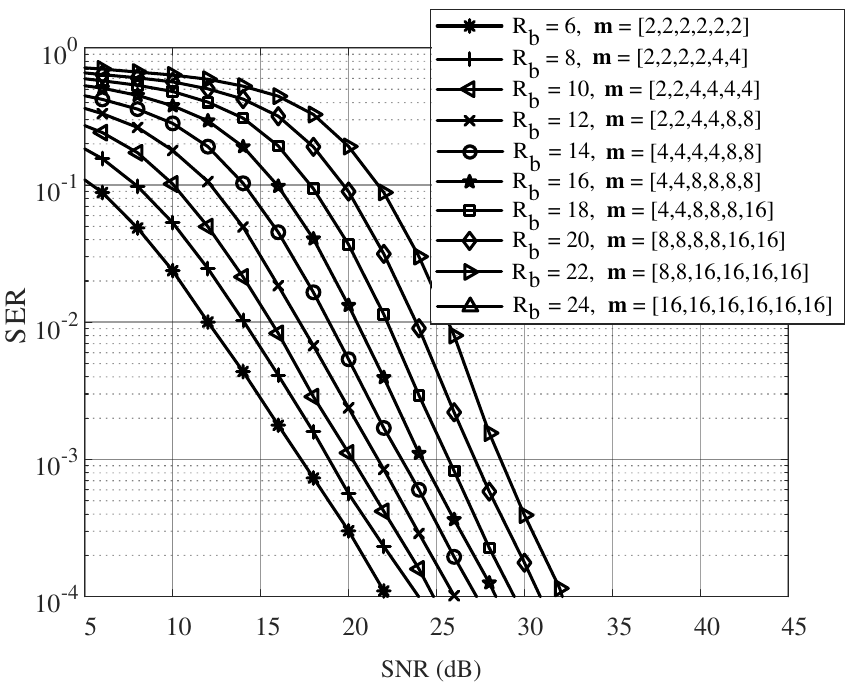}
		\caption{$R_b = 6,8,\ldots, $ and $24 $   }
	\end{subfigure}
	\caption{SER performance of representative  VMMs with  the same distance.}
	\label{SER_eq2}
\end{figure*}

\vspace{-0.2cm}
\subsection{SER Performance: Users With Equal Distance}

\label{eqd}

 \textit{1) Same modulation order case:}
We first start  by evaluating the SER performance of VM-SCMA with  the same modulation order  and  assume  $\mathbf d =  \mathbf 1^{J \times 1}$, which is generally considered in existing SCMA codebook designs.  The main representative codebooks  for comparison with the proposed ones are the     Chen's codebook \cite{chen2020design},    StarQAM codebook \cite{yu2015optimized} and LDS codebook. The basic constellations in Fig. \ref{basic} are employed to construct the LDS codebooks, which can be viewed as the proposed design without permutation.  As can been seen in Fig. \ref{SER_Meq},  the proposed codebooks achieve the same  or better SER performance compared to the benchmarks  for $M=2,4,8$ and $16$.  This indicates the proposed MCs are all well optimized, which is vitally  important when applying these MCs to  VM-SCMA and AVM-SCMA.

 \textit{2) Diverse modulation orders case:} Next, we evaluate the performance of the VM-SCMA when diverse modulation orders are supported in a VMM. 
Fig. \ref{SER_eq}  presents the SER performance of the proposed VM-SCMA for various modulation order combinations. Fig. \ref{SER_eq}(a) shows the $J$ users' SER performance  of $\mathbf m = [2, 4, 8, 8, 16, 16]$ and $R_b = 17$ bits, where four different modulation orders are supported.    As can be seen from the figure, all users achieve approximately the same SER performance in the medium-to-high SNR range, even though the modulation orders are different, due to the well-optimised VMM and the power allocation scheme based on the proposed AIPD criteria.   

Fig. \ref{SER_eq}(b) presents the SER performance  of the VMMs   considered in Example 1. 
Clearly, the optimized VMM with   $\mathbf m = [2, 4, 8, 8, 16, 16]$ (${{\mathbf{M}}_{4\times 6}^{(3)}}$) achieves better performance than that   of    $\mathbf m = [2, 16, 8, 16, 2, 16]$  (${{\mathbf{M}}_{4\times 6}^{(4)}}$). 
In Subsection \ref{dvd}, we will show that ${{\mathbf{M}}_{4\times 6}^{(5)}}$   achieves the best error rate performance for {$ \mathbf d = [4.70, 4.60,1.62, 1.25, 1.20, 1.13]$.  

Fig. \ref{SER_eq}(c)  compares the SER performance of the proposed VM-SCMA with diverse modulations and VM-SCMA with the same modulation order    for $R_b =12$ and  $R_b =18$. Note that the proposed  codebooks with   the same modulation order are  employed for comparison as they achieve  the same or better SER performance compared to the benchmarch codebooks.   The proposed VM codebooks  with $\mathbf m = [4, 4, 8, 8, 16, 16]$ and  $\mathbf m = [2, 2, 4, 4, 8, 8]$  can not only  support three different modulation orders, but  achieve  the approximately the same SER performance with that of the same  modulation cases.

 The overall data rate   can be supported by the proposed VM-SCMA is $R_b \in  \{6,7,\ldots,24\}$. In contrast, the overall data rate of the existing sparse codebooks for SCMA is restricted  to $R_b = J\log_2(M), M =2, 4, 8, 16$.  
 We further present the SER performance of  VM-SCMA with  a representative modulation combination  for each $R_b$.   For simplicity, the SER curves are presented in two  figures,  as shown in Fig. (\ref{SER_eq2}). As can be  seen, the proposed  VM-SCMA can support diverse data rates  and achieve  promising SER performance.  It is also worth mentioning that  in addition to the sparse codebooks presented in Fig. (\ref{SER_eq}) and Fig. (\ref{SER_eq2}),  the proposed VM-SCMA can be employed to design the sparse codebooks with  arbitrary modulation order combinations, which provides significant flexibility for SCMA systems.

\vspace{-0.3cm}
\subsection{SER Performance: Users With Diverse Distances}
\label{dvd}

\begin{figure}
	\centering
	\begin{subfigure}{0.24 \textwidth}
  \includegraphics[width=1 \textwidth]{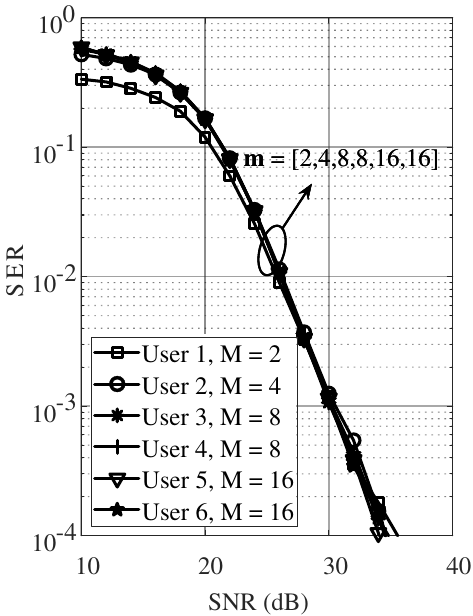}
		\caption{Single user SER performance. }
	\end{subfigure}
	\begin{subfigure}{0.24\textwidth}
  \includegraphics[width= 1 \textwidth]{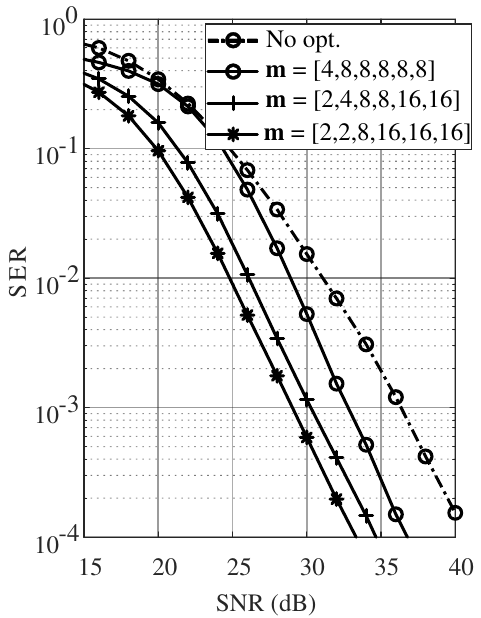}
		\caption{ SER performance.  }
	\end{subfigure}
	\caption{$R_b = 17$ bits.}
	\label{SER_Rb17}
\end{figure}

\begin{figure}
	\centering
	\begin{subfigure}{0.24 \textwidth}
  \includegraphics[width=1 \textwidth]{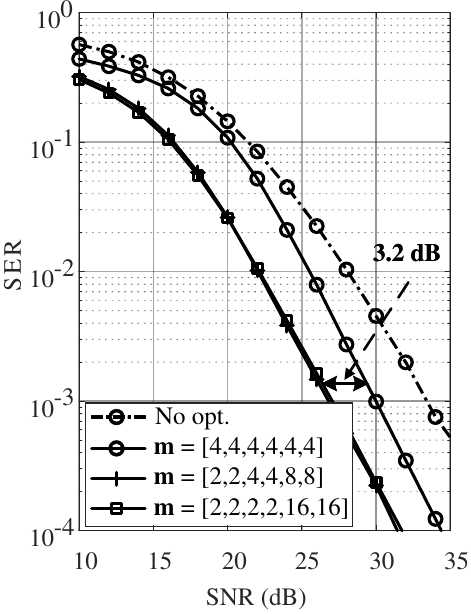}
		\caption{ SER performance. }
	\end{subfigure}
	\begin{subfigure}{0.24\textwidth}
  \includegraphics[width=1 \textwidth]{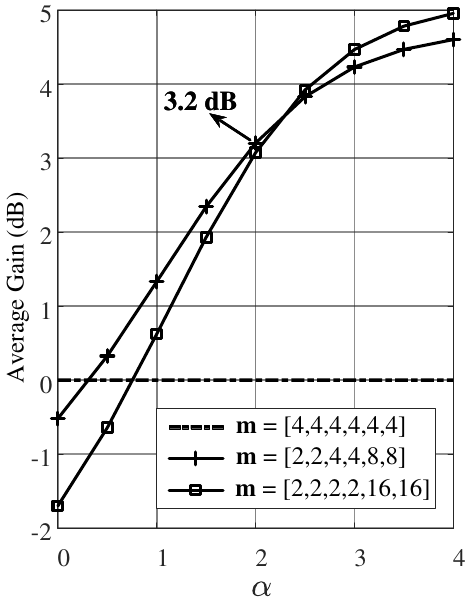}
		\caption{  Analytical gain ($G$).  }
	\end{subfigure}
	\caption{$R_b = 12$ bits.}
	\label{SER_Rb12}
\end{figure}

\begin{figure}
	\centering
	\begin{subfigure}{0.24 \textwidth}
  \includegraphics[width=1 \textwidth]{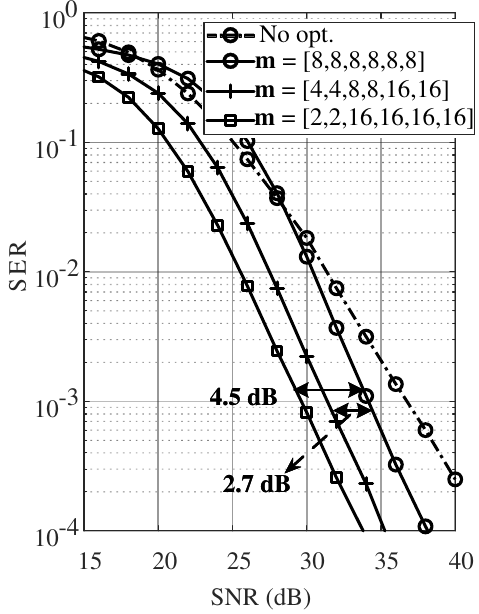}
		\caption{ SER performance. }
	\end{subfigure}
	\begin{subfigure}{0.24\textwidth}
  \includegraphics[width= 1 \textwidth]{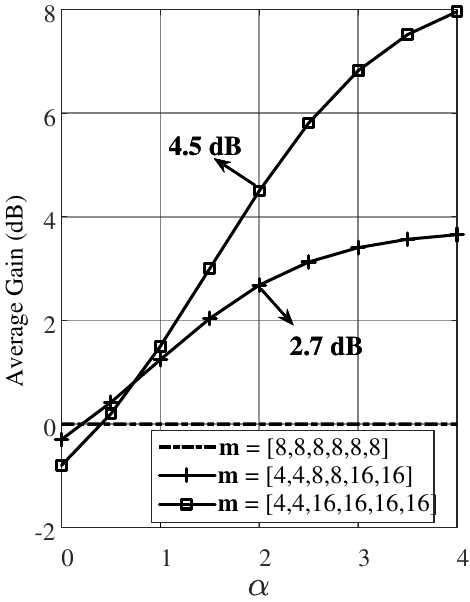}
		\caption{  Analytical gain ($G$).  }
	\end{subfigure}
	\caption{$R_b = 18$ bits.}
	\label{SER_Rb18}
\end{figure}
\begin{figure}
	\centering
	\begin{subfigure}{0.24 \textwidth}
  \includegraphics[width=1 \textwidth]{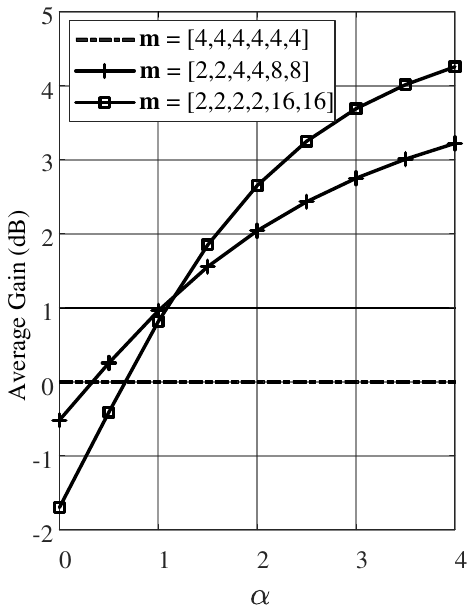}
		\caption{$R_b =12$ bits.}
	\end{subfigure}
	\begin{subfigure}{0.24\textwidth}
  \includegraphics[width= 1 \textwidth]{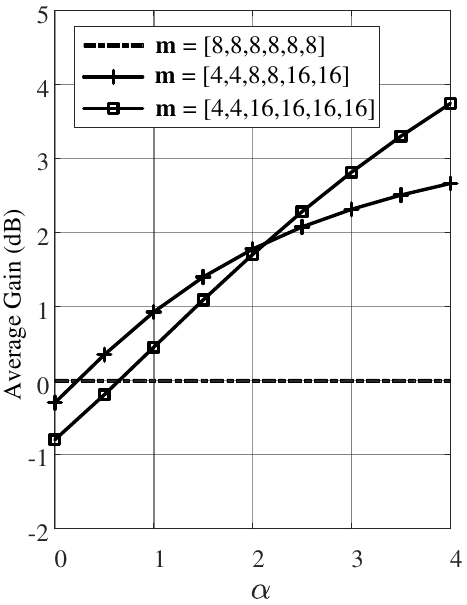}
		\caption{ $R_b =18$ bits.  }
	\end{subfigure}
	\caption{Averaged analytical gain over a cell.}
	\label{Gain_ave}
    \vspace{-0.3cm}
\end{figure}
We now consider the case where the users  are   at different locations in a cell. In particular, $\mathbf d = [4.70, 4.60,1.62, 1.25, 1.20, 1.13]$ and $\alpha =2$ are first  employed to evaluate the   performance of the proposed VM-SCMA systems.  Fig. \ref{SER_Rb17} presents the SER performance of the VMMs presented in Example 1.  As can be  seen from the figure, the SER performance for the users with different  modulation orders  can be guaranteed to be same in the medium-to-high SNR range.  Unlike the case in  Fig. \ref{SER_eq},   the codebook with modulation order $\mathbf m =[2, 2, 8, 16, 16, 16]$ (${{\mathbf{M}}_{4\times 6}^{(5)}}$) achieves the best SER performance when  $\mathbf d = [4.70, 4.60,1.62, 1.25, 1.20, 1.13]$.

 Fig. \ref{SER_Rb12} and  Fig. \ref{SER_Rb18} show the SER performance of the proposed VM codebooks with diverse modulation orders and the codebook with same modulation order of $R_b =12$ bits and $R_b= 18$ bits, respectively. Specifically, the VMMs that achieve optimal VM-SCMA  are employed.  In Fig. \ref{SER_Rb12}(a), the proposed codebooks with $\mathbf m =[4,4,4,4,4,4], \mathbf m =[2,2,4,4,8,8]$ and $ \mathbf m =[2,2,2,2,16,16]$ achieve the same rate of $R_b = 12$ bits.  It is noted that the proposed VM codebooks  with $  \mathbf m =[2,2,4,4,8,8]$ and $ \mathbf m =[2,2,2,2,16,16]$   achieve about $3.2$ dB gain over that of the codebook with the same modulation order. The dash lines in Fig. \ref{SER_Rb12}(a) and Fig. \ref{SER_Rb18}(a), i.e.  the curves ``No opt.'', denote the SER performance of the worst user of the     the same modulation order codebook  without the proposed power allocation scheme. Fig. \ref{SER_Rb12}(b) also shows the analytical gain over  different path loss exponents ($\alpha$) using the Remark $3$ in Subsection \ref{adVM}. It can be seen    that  the gain becomes more prominent as the   $\alpha$ increases. Similar results are also observed for the proposed  codebooks with $R_b =18$ bits, whose  performance is shown in Fig.    \ref{SER_Rb18}.  It shows that the codebook  with $  \mathbf m =[2,2,2,2,16,16]$ achieves $4.5$ dB gain over the codebook with the same modulation order of $\alpha =2$, and  almost $8$ dB gain can be achieved with   $\alpha =4$.

 Fig. \ref{Gain_ave} shows the average gain of the proposed VM-SCMA with the codebook of different modulation orders over the codebook of same modulation order, where $R_b =12$ bits and $R_b =18$ bits are considered. The average gain are obtained by averaging over $N_s = 1000$  randomly generated distance vectors over the cell. As can be seen from the figure, the gain between the proposed    VM-SCMA with different modulation orders and that   with same modulation order becomes larger as $\alpha$ increases. In addition, for different $\alpha$, the optimal VMM  for a given rate of a cell may also be different.  This indicates that the VMM should be selected to adapt to the users' channel conditions.

 \begin{figure}[tbp]
	\centering
	\begin{subfigure}{0.48 \textwidth}
  \includegraphics[width=1 \textwidth]{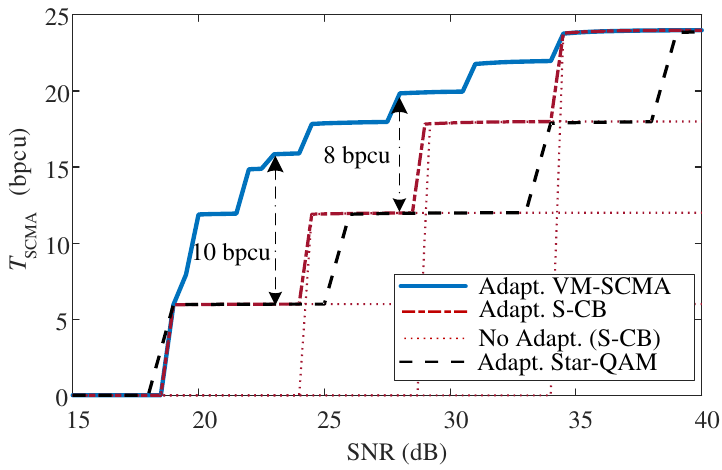}
		\caption{ Effective throughput with fixed distance. }
	\end{subfigure}
	\begin{subfigure}{0.48\textwidth}
  \includegraphics[width=1 \textwidth]{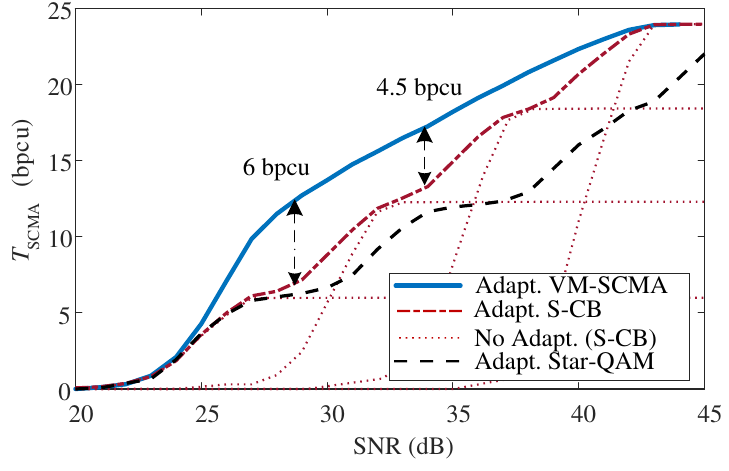}
		\caption{Average  effective throughput.  }
	\end{subfigure}
 	\caption{Effective throughput performance.}
	\label{ETR}
    \vspace{-0.3cm}
\end{figure}

 \subsection{Effective Throughput of AVM-SCMA}
 \label{etr}
 This subsection evaluates  the effective throughput performance of the proposed   AVM-SCMA system.  Specifically,  we   present the effective throughput of $\mathbf d = [4.70, 4.60,1.62, 1.25, 1.20, 1.13]$ and the average effective  throughput with  $\text{SER}_\text{th} = 0.01$   in Figs.  \ref{ETR}(a)  and  Fig.  \ref{ETR}(b), respectively.  Similar to Fig. \ref{Gain_ave},   the average effective  throughput  is obtained by averaging over $N_s=1000$ distance vectors over the cell. The   curves ``AVM-SCMA'' and  ``Adapt. S-CB''  are the proposed adaptive scheme with the VM codebooks given in Table \ref{VMMM} and the proposed codebooks with the same modulation order, respectively. In addition, we also evaluate the average effective  throughput with the StarQAM codebooks, which is the ``Adapt. StarQAM''.   The main observations of Fig. \ref{ETR}  are summarized as follows:
 
\begin{itemize}
    \item The proposed   AVM-SCMA achieves significant effective  throughput   improvements over the benchmarks. In particular,  about  $10$ bits gain can be achieved for $\mathbf d = [4.70, 4.60,1.62, 1.25, $ $1.20, 1.13]$  at  $\text{SNR} = 23$ dB, when compared to ``Adapt. S-CB''.  This indicates that the proposed AVM-SCMA can transmit an additional 10 bits while maintaining a SER constraint of $\text{SER}_\text{th} = 0.01$.
    
    \item As can been seen from  Fig.  \ref{ETR}(b), the proposed AVM-SCMA   achieves about 
 $6$ bits and  $4.5$ bits gains over that of ``Adapt. S-CB''   at  $\text{SNR} = 28$ dB and    $\text{SNR} = 34$ dB, respectively.  It is worth noting that the gain becomes   more prominent when compared to adaptive modulation with the StarQAM codebook.
\end{itemize}

\section{Conclusion}
 
In this paper, we have introduced a novel  VM-SCMA system that enables SCMA users to employ the codebooks
with diverse modulation orders. To guide the VM-SCMA design,  we   established the design criteria of VM-SCMA  based on the proposed    AIPD  and the asymptotic upper bound of sum-rate. In addition, we   considered a more generalized case where SCMA users are randomly distributed within a cell.  The detailed design of MCs, power allocation, codebook allocation and  VMM have been  investigated.  To further improve the spectrum   efficiency,  we     proposed an AVM-SCMA scheme that adaptively selects VM codebook according to   users' statistical SNRs by maximizing the effective throughput of the VM-SCMA systems subject to a reliable error rate constraint.  Numerical results demonstrated the benefits and superiority
of the proposed VM-SCMA and adaptive VM-SCMA in terms of flexibility of providing diverse data rates,  symbol error rate performance improvement and    
throughput performance enhancements over the state-of-the-art SCMA codebooks.

\appendices
\section{The proposed MC pool}
\label{MCpOOL}

The proposed MC pool  is presented. Alternatively, one can find these MCs and the proposed   VM codebooks at:   \url{https://github.com/ethanlq/SCMA-codebook}. The AIPD values of $ \boldsymbol {\mathcal C}_{\text{MC}}^{M=2}$, $ \boldsymbol {\mathcal C}_{\text{MC}}^{M=4}$, $ \boldsymbol {\mathcal C}_{\text{MC}}^{M=8}$ and $ \boldsymbol {\mathcal C}_{\text{MC}}^{M=16}$ are given as $0.25,  2,   9.9$ and $ 39$, respectively.
       \begin{equation}
  \setlength{\arraycolsep}{2.0pt} 
  \footnotesize   
     \boldsymbol {\mathcal C}_{\text{MC}}^{M=2}=
\left[\begin{matrix}
  -0.707 & 0.707 \\
  -0.707 & 0.707 \\
      \end{matrix}\right],  \notag
        \end{equation}  

\vspace{-1em} 
      \begin{equation}
  \setlength{\arraycolsep}{2.0pt} 
  \footnotesize   
     \boldsymbol {\mathcal C}_{\text{MC}}^{M=4}=
\left[\begin{matrix}
   0.50+0.50i &  0.50-0.50i & -0.50+0.50i  &-0.50-0.50i \\
   0.50-0.50i & -0.50+0.50i & -0.50-0.50i  & 0.50+0.50i
      \end{matrix}\right],  \notag
        \end{equation}  
The codewords corresponding to the $1$st to $4$th positions, and $5$th to $8$th positions in  $\boldsymbol{\mathcal{C}}_{\text{MC}}^{M=8}$  are respectively given as follows:
\vspace{-0.3em} 
      \begin{equation}
  \setlength{\arraycolsep}{2.0pt} 
  \footnotesize   
\left[\begin{matrix}
-0.67-0.58i& 0.58i	&-0.33&	0.58i\\
0.33 &	-0.67-0.58i	&-0.67+0.58i&0.67+0.58i
      \end{matrix}\right],  \notag
        \end{equation}  
         \vspace{-1em} 
      \begin{equation}
  \setlength{\arraycolsep}{2.0pt} 
  \footnotesize   
\left[\begin{matrix}
-0.67+0.58i&  0.33& 0.67-0.58i & 0.67+0.58i 		\\
-0.58i & 0.67-0.58i& 0.58i	&-0.33
      \end{matrix}\right]. \notag
        \end{equation}  
        
The codewords corresponding to the $1$st to $4$th positions,  $5$th to $8$th positions, $9$th to $12$th positions and $13$th to $16$th positions  in  $\boldsymbol{\mathcal{C}}_{\text{MC}}^{M=16}$  are respectively given as follows:
 \vspace{-0.3em} 
      \begin{equation}
  \setlength{\arraycolsep}{2.0pt} 
  \footnotesize   
\left[\begin{matrix}
0.67+0.67i&	0.67 - 0.22i	&0.67+ 0.22i	&0.67  - 0.67i\\
0.22- 0.22i &	0.67 - 0.22i	&-0.67 - 0.22i	&-0.22 -0.22i\\
      \end{matrix}\right],  \notag
        \end{equation}  
         \vspace{-1em} 
      \begin{equation}
  \setlength{\arraycolsep}{2.0pt} 
  \footnotesize   
\left[\begin{matrix}
-0.22+ 0.67i	&-0.22 - 0.22i&	-0.22+ 0.22i&	-0.22- 0.67i\\
0.22 - 0.67i	&0.67 - 0.67i&	-0.67- 0.67i&	-0.22 - 0.67i
      \end{matrix}\right], \notag
        \end{equation} 
         \vspace{-1em} 
      \begin{equation}
  \setlength{\arraycolsep}{2.0pt} 
  \footnotesize   
\left[\begin{matrix}
0.22 + 0.67i&	0.22 - 0.22i&	0.22 + 0.22i&	0.22 - 0.67i\\
0.22+ 0.67i	 &  0.67 + 0.67i	&-0.67+ 0.67i	&-0.22 + 0.67i
      \end{matrix}\right],\notag
        \end{equation} 
                 \vspace{-1em} 
      \begin{equation}
  \setlength{\arraycolsep}{2.0pt} 
  \footnotesize   
\left[\begin{matrix}
-0.67 + 0.67i&	-0.67 - 0.22i&	-0.67 + 0.22i	&-0.67 - 0.67i\\
0.22+ 0.22i	&0.67 + 0.22i	&-0.67 + 0.22i&	-0.22+ 0.22i
      \end{matrix}\right]. \notag
        \end{equation}


\ifCLASSOPTIONcaptionsoff
  \newpage
\fi



%
\bibliography{ref} 
\bibliographystyle{IEEEtran}

\end{document}